\documentclass[preprint,11pt,nofootinbib]{revtex4-1}

\usepackage{amsmath}
\usepackage{amsfonts}
\usepackage{amssymb}
\usepackage{mathtools}
\usepackage{graphicx}
\usepackage{eucal}
\usepackage{tikz}
\usepackage{tikz-feynman}
\usepackage{mathtools}
\usepackage{braket}
\usepackage{slashed}
\usepackage{appendix}
\usepackage{hyperref}
\usepackage{enumitem}
\usepackage{listings}
\usepackage{multirow}
\usepackage{lipsum}
\usepackage{array}
\usepackage{bm}

\usepackage{color}
\usepackage{comment}

\newcommand{\vecmat}[2]{\begin{pmatrix}
#1\\
#2
\end{pmatrix}}

\newcommand{\barv}[1]{\overline{#1}}

\newcolumntype{P}[1]{>{\centering\arraybackslash}p{#1}}

\begin{document}

\begin{flushright}

KANAZAWA-23-02

OU-HET 1166
\end{flushright}

\title{\Large{Probing chirality structure in lepton-flavor-violating Higgs decay  $h\to\tau\mu$ at the LHC}}

\author{Mayumi Aoki$^1$}
\email{mayumi.aoki@staff.kanazawa-u.ac.jp}
\author{Shinya Kanemura$^2$}
\email{kanemu@het.phys.sci.osaka-u.ac.jp}
\author{Michihisa Takeuchi$^{34}$}
\email{takeuchi@mail.sysu.edu.cn}
\author{Lalu Zamakhsyari$^1$}
\email{lalu\_zamakhsyari@stu.kanazawa-u.ac.jp}

\affiliation{$^1$Institute for Theoretical Physics, Kanazawa University, Kanazawa 920-1192, Japan\\
$^2$Department of Physics, Osaka University, Toyonaka, Osaka 560-0043, Japan\\
$^3$School of Physics and Astronomy, Sun Yat-sen University, 519082 Zhuhai, China\\
$^4$Graduate School of Information Science and Technology, Osaka University, Suita, Osaka 565-0871, Japan}

\begin{abstract}
A phenomenological study for determining the chirality structure in lepton-flavor-violating Higgs (hLFV) decays $h \to \tau\mu$ at the LHC
is presented.
We estimate the effects of the $\tau$ polarization in the analysis and
the importance of determining the relative visible momentum ratio $x$, and show the analysis 
with a collinear mass $m_{col1}$ by assuming one missing particle is appropriate.
We find that the sensitivity would be generically affected up to $\pm$ $4-6$~\% in terms of the BR$(h\to \tau\mu)$ upper bound, 
and show the altered bounds on the $(|y_{\mu\tau}|, |y_{\tau\mu}|)$ plane.
We further study the benchmark scenarios, 
and demonstrate the sensitivity study for the chirality structure using the relative visible momentum ratio.
We find that the two fully polarized cases, the $\tau_R$ and $\tau_L$ scenarios consistent with the recently reported excess,
are distinguishable at 2$\sigma$ level for 1000~fb$^{-1}$. We also show that
a further improved study potentially provides a similar sensitivity already for 139~fb$^{-1}$.
\end{abstract}

\maketitle

\clearpage
\section{Introduction}
The discovery of 125 GeV Higgs boson~\cite{CMS:2012qbp, ATLAS:2012yve} is certainly
one of the most important discoveries in particle physics. 
It has been proven so far that 
properties of the Higgs boson are consistent with the standard model (SM) predictions.
However, the present data do not rule out possibilities of new mechanisms 
of electroweak symmetry breaking or new electroweak physics beyond the SM.
The effects of such new physics would modify the couplings of the 125 GeV Higgs boson 
from the SM predictions or introduce new interactions that are absent in the SM.
An example of the latter is the lepton-flavor violating Higgs (hLFV) processes,
which involve introduction of off-diagonal components of the Yukawa coupling $y_{ij}$ for the lepton sector 
in the effective Lagrangian term:
\begin{equation}
\label{lageq}
-\mathcal{L}_{hll}=y_{ij}
\bar{l}_{Li}hl_{Rj}+h.c.,
\end{equation}
where the hLFV couplings are induced at the tree level or as the loop-effect. 
The two Higgs doublet models (2HDMs), for instance, induce the hLFV couplings at the tree level~\cite{Diaz-Cruz:1999sns, Kanemura:2005hr, Crivellin:2015mga, Crivellin:2015hha, Botella:2015hoa, Omura:2015xcg,  Sher:2016rhh, Herrero-Garcia:2017xdu, Hou:2019grj, Nomura:2019dhw, Vicente:2019ykr,  Hou:2020tgl},
meanwhile the seesaw models \cite{Pilaftsis:1992st, Arganda:2004bz, Arganda:2014dta, Aoki:2016wyl, Arganda:2016zvc,  Thao:2017qtn, Marcano:2019rmk}  and minimal supersymmetric standard models~\cite{Diaz-Cruz:1999sns, Diaz-Cruz:2002ezb, Brignole:2003iv, Arganda:2004bz,  Kanemura:2004cn, Crivellin:2010er,  Giang:2012vs, Arhrib:2012mg, Arana-Catania:2013xma,  Abada:2014kba, Arganda:2015naa, Arganda:2015uca,  Zhang:2015csm, Fathy:2016vli,  Gomez:2017dhl} at the one-loop level.

The off-diagonal components are responsible 
 for hLFV decays,
$h\to l_il_j$ $(i\neq j)$, 
where $h\to l_i l_j$ mean the sum of the processes $h\to l_i^+ l_j^-$ and $h\to l_i^- l_j^+$. 
Any observation of hLFV decays is a clear evidence of new physics beyond the SM. 
The prospects of probing hLFV decays at the LHC and future colliders have been widely explored 
in a model independent way~\cite{Diaz-Cruz:1999sns, Cotti:2005ah, Goudelis:2011un, Blankenburg:2012ex, Harnik:2012pb, Celis:2013xja, Banerjee:2016foh, Chakraborty:2017tyb, Davidek:2020gbw, Barman:2022iwj}
and in the various models~\cite{
Diaz-Cruz:1999sns, Kanemura:2005hr, Crivellin:2015mga, Crivellin:2015hha, Botella:2015hoa, Omura:2015xcg,  Sher:2016rhh, Herrero-Garcia:2017xdu, Hou:2019grj, Nomura:2019dhw, Vicente:2019ykr,  Hou:2020tgl,
Pilaftsis:1992st, Arganda:2004bz, Arganda:2014dta, Aoki:2016wyl, Arganda:2016zvc,  Thao:2017qtn, Marcano:2019rmk,
Diaz-Cruz:1999sns, Diaz-Cruz:2002ezb, Brignole:2003iv, Arganda:2004bz,  Kanemura:2004cn, Crivellin:2010er,  Giang:2012vs, Arhrib:2012mg, Arana-Catania:2013xma,  Abada:2014kba, Arganda:2015naa, Arganda:2015uca,  Zhang:2015csm, Fathy:2016vli,  Gomez:2017dhl,
Davidson:2012ds, Bressler:2014jta, Dery:2014kxa, Heeck:2014qea, deLima:2015pqa, He:2015rqa, Cheung:2015yga, Baek:2015mea,  Baek:2015fma, Hue:2015fbb, Chang:2016ave,  Chakraborty:2016gff, Lami:2016mjf, Altmannshofer:2016oaq,  Herrero-Garcia:2016uab, Hayreter:2016aex, Fonseca:2016xsy, Qin:2017aju, Hong:2020qxc,  Nguyen:2020ehj, Hung:2021fzb, Zhang:2021nzv,  Zeleny-Mora:2021tym,  Hundi:2022iva, Hung:2022kmv, Abada:2022asx}.

Among $h\to l_il_j$ processes,  
the $h\to\mu e$ is strongly suppressed due to 
the stringent constraint on 
$y_{\mu e}$ from the rare $\mu$ decay \cite{MEG:2016leq}.
On the other hand, the $h\to\tau e$ and $h\to\tau\mu$ are less constrained.
In this paper, we focus on $h\to\tau\mu$ since we would expect naturally larger effects 
and also $h\to\tau e$ is known to be experimentally more challenging.
The relevant Yukawa coupling $y_{\tau\mu}$ is also constrained by the low-energy lepton-flavor violating (LFV) processes,
where the strongest one is coming from $\tau\to\mu\gamma$. 
However, those constraints are
still weaker than the current bounds in the hLFV decays by the ATLAS  \cite{ATLAS:2019pmk, ATLAS:2022conf} and CMS \cite{CMS:2021rsq} collaborations. 
Even with the future $\tau\to\mu\gamma$ measurement 
at Belle II \cite{Belle-II:2018jsg}, the constraints on $y_{\tau\mu}$ from 
the hLFV decays will still be the most stringent.
Current measurements of branching ratios (BRs) of the hLFV decays $h\to l_il_j$ 
at the LHC give the constraints on $\bar{y}_{ij}\equiv \sqrt{|y_{ij}|^2+|y_{ji}|^2}$.
At the future hadron and lepton colliders, 
we expect to improve the limit on $\bar{y}_{\tau\mu}$
and also to probe the chirality structure of the Yukawa matrix, 
for example, the ratio $y_{\tau\mu}/y_{\mu\tau}$ through the measurements 
in the hLFV decays~\footnote{There are models to predict such an asymmetry. 
See e.g., \cite{Peccei:1986pn, Chen:2010su, Chiang:2015cba, Chiang:2017fjr, Chiang:2018bnu}.}.
Therefore, the way to probe the chirality structure in the leptonic Yukawa sector would 
play an important role in distinguishing the models if such processes are observed.

In this paper, 
we study the hLFV decay $h\to \tau\mu$ at the LHC, and study the chirality structure of the process in the SMEFT.
The numbers of signal events for $h\to \tau_L\mu_R$ and $h\to \tau_R\mu_L$, and those of backgrounds are evaluated.
We focus on the gluon fusion (ggF) Higgs boson production in the simulation as a main production mode and 
hadronic $\tau$ decay modes are considered.
We show that the collinear mass based on one missing particle assumption is important 
for getting the better discrimination sensitivity between $h\to \tau_L\mu_R$ and $h\to \tau_R\mu_L$.
We show the resulting sensitivity on the off-diagonal elements of the Yukawa coupling matrix, 
and express it in the ($|y_{\tau\mu}|$,  $|y_{\mu\tau}|$) plane.

This paper is organized as follows. 
In Sec.\,\ref{exp},
we summarize the current constraints on the $\mu-\tau$ sector.
Theoretical frameworks of the hLFV for new physics beyond the SM are described in Sec.\,\ref{theo}.
In Sec.\,\ref{sec:analysis}, we discuss the collider simulation of the hLFV process $h\to\tau\mu$ 
taking the polarization effects of $\tau$ lepton into account,
and present the results on the upper limit of the branching ratio (BR) of $h\to\tau\mu$, and the corresponding hLFV Yukawa couplings $y_{\tau\mu}$ and $y_{\mu\tau}$.
We also consider the three benchmark scenarios with different chirality structures, 
which are inspired by the recent excess reported by ATLAS, 
and demonstrate the way to discriminate the scenarios.
Sec.\,\ref{sec:conclusion} is devoted to the discussion and summary.


\section{Experimental Status }
\label{exp}

The ATLAS and CMS have searched for $h\to l_i l_j$ and provide 
the upper bounds on those BRs.
For our interest in hLFV $h\to\tau\mu$, the upper limit on the BR$(h\to \tau\mu)$ at 95~\%~C.L. is reported as
\cite{CMS:2021rsq, ATLAS:2019pmk}
\begin{equation}
{\rm BR}(h\to\tau\mu) \le  0.28\text{ \%} \ \text{(ATLAS)} \ \ {\rm and}\ \  0.15\text{ \%} \ \text{(CMS)},
\end{equation}
at the total integrated luminosity of 36.1 and 139 fb$^{-1}$, respectively\footnote{Recently, 
the ATLAS collaboration gives the bound BR($h\to\tau\mu$) $<$ 0.18 \%, 
which is obtained from the combined searches in $\mu\tau$ and $e\tau$ channels  with 138 fb$^{-1}$, and
exhibits 2.2 $\sigma$ level upward deviation in comparison with the expected sensitivity 0.09~\%~\cite{ATLAS:2022conf}.
}.
These limits are interpreted as $\sqrt{|y_{\mu\tau}|^2+|y_{\tau\mu}|^2} < 1.5 \times 10^{-3}$(ATLAS) and  
$1.11\times 10^{-3}$(CMS), respectively.
The projected limit at the high luminosity LHC (HL-LHC, 3000 fb$^{-1}$) has been estimated at 
 $\sim 10^{-4}$~\cite{Barman:2022iwj,Davidek:2020gbw}.
 Furthermore, the hLFV decay would also be searched for in the future $e^+e^-$ colliders, 
 where the sensitivity for the hLFV branching ratios would also reach $\sim10^{-4}$
 as shown by several analyses~\cite{Kanemura:2005hr, Chakraborty:2016gff, Chakraborty:2017tyb, Qin:2017aju, Li:2018cod}.
 
\begin{table}[t]
\centering
\begin{tabular}{c| c |c}
\hline
LFV process & Present bound BR & Future sensitivity\\
\hline
$\tau\to \mu\gamma$ & $4.4\times 10^{-8}$ \cite{BaBar:2009hkt} &  $\sim 10^{-9}$ \cite{Belle-II:2018jsg}\\
$\tau\to \mu\mu\mu$ & $2.1\times 10^{-8}$ \cite{Hayasaka:2010np}&  $5\times 10^{-10}$ \cite{Belle-II:2018jsg}\\
$\tau^-\to e^-\mu^+\mu^-$ & $2.7\times 10^{-8}$ \cite{Hayasaka:2010np}& $5\times 10^{-10}$ \cite{Belle-II:2018jsg}\\
$\tau^-\to \mu^-e^+e^-$ & $1.8\times 10^{-8}$ \cite{Hayasaka:2010np}& $5\times 10^{-10}$ \cite{Belle-II:2018jsg}\\
$\tau^-\to e^+\mu^-\mu^-$ & $1.7\times 10^{-8}$ \cite{Hayasaka:2010np}&$4\times 10^{-10}$ \cite{Belle-II:2018jsg}\\
$\tau^-\to \mu^+e^-e^-$ & $1.5\times 10^{-8}$ \cite{Hayasaka:2010np}&$3\times 10^{-10}$ \cite{Belle-II:2018jsg}\\
\hline
\end{tabular}
\caption{Current experimental bounds and future sensitivities for several low-energy LFV observables in the $\tau$-$\mu$ sector.}
\label{lfv_table}
\end{table}

The LFV Yukawa couplings relevant to the $h\to l_i l_j$ process also induce the low-level LFV processes, such as 
$l_i\to l_j\gamma$ and $l_i\to l_jl_k l_k$ processes. 
The bounds for the relevant low-level LFV processes in the $\tau$-$\mu$ sector are given in Table~\ref{lfv_table}. 
Although they are all around $\sim 10^{-8}$, the $\tau\to \mu\gamma$ measurement provides the strongest bound, 
$\sqrt{|y_{\mu\tau}|^2+|y_{\tau\mu}|^2} < 0.016$~\cite{ BaBar:2009hkt, Harnik:2012pb}.
This bound is still weaker than the current bound from the hLFV decay process of $h\to \tau \mu$.
All future sensitivities are promised to increase up to one or two orders of magnitude as summarized in Table \ref{lfv_table}. 
The hLFV couplings of $y_{\mu\tau}$ and $y_{\tau\mu}$ are also constrained from the measurements of 
the anomalous magnetic moment $(g-2)_\mu$~\cite{Abi:2021_et_al}, 
the muon electric dipole moment (EDM)~\cite{Bennett:2009_et_al},
the tau EDM~\cite{ARGUS:2000, Belle:2003}, 
as well as the lepton-nucleus scattering $\mu N \to \tau X$~\cite{Kanemura:2004jt, Takeuchi:2017btl, Kiyo:2021ibt},
whereas they are weaker than the constraints from $\tau \to \mu\gamma$.

\section{Theoretical Framework for the hLFV process}
\label{theo}
We here briefly introduce the hLFV couplings in a model-independent manner 
following the effective-field theory extension of the SM (SMEFT).
This is well motivated in the current situation with the absence of new physics signatures at the LHC, 
which supports any new particles responsible for a new physics would be well beyond the current electroweak scale. 
The type III 2HDM is also introduced as an example concrete model which induces the hLFV couplings at tree level.

\subsection{Standard model effective-field theory}

The general form of the SMEFT is given by 
\begin{equation}
\mathcal{L}_{EFT}=\mathcal{L}_{SM}^{(4)}+\frac{1}{\Lambda}\sum_{a}C_a^{(5)}Q_a^{(5)}+\frac{1}{\Lambda^2}\sum_{a}C_a^{(6)}Q_a^{(6)}+\mathcal{O}
\left(\frac{1}{\Lambda^3}\right)\,,
\end{equation}
where $\Lambda$ is the NP scale, $Q^{(d)}_a$ are the $d$-dimension operators 
composed of the SM fields, and their associated Wilson coefficients $C_a^{(d)}$ are in general complex
(flavor indices have been suppressed).
The dimension 5 operator is the Weinberg operator that gives rise to 
the neutrino Majorana mass~\cite{Dedes:2017zog}, so this is not our concern. The higher dimension $d>6$ 
contributions are suppressed for hLFV, so that our main focus for hLFV is the dimension 6 operators. 
The operator $Q_a^{(6)}$ is usually presented in the Warsaw basis~\cite{Grzadkowski:2010es, Dedes:2017zog}, 
and the one that is relevant for hLFV is given by 
\begin{equation}
\label{qhlfv}
Q^{l\varphi}= (\varphi^{\dagger}\varphi)\barv{L}_L'C^{l\varphi}l_R'\varphi\,,
\end{equation}
where 
$L_L'$ and $l_R'$ are the lepton SU(2)$_L$-doublet and singlet, respectively, 
in flavor space, and $\varphi$ is the SU(2)$_L$ Higgs doublet.
Other dimension-six Warsaw operators that induce the hLFV couplings are suppressed, 
or can be reduced to Eq.\eqref{qhlfv} via field redefinitions, Fierz identities and equations of motions \cite{Herrero-Garcia:2016uab}. 

After the  electroweak symmetry breaking,
the lepton Yukawa Lagrangian in the SM,
$-\mathcal{L}^{SM}_{Y}=\bar{L}_L'Y^ll_R'\varphi+h.c.$, where $Y^l$ is the Yukawa matrix,
and the effective low energy Lagrangian
by the operator $Q^{l\varphi}$, $\mathcal{L}_{hLFV}$, can be written as
\begin{align}
-(\mathcal{L}_{Y}^{SM}
+\mathcal{L}_{hLFV})
&= 
\barv{l}_L'\left[\frac{v}{\sqrt{2}}\left(Y^l-\frac{C^{l\varphi}v^2}{2\Lambda^2}\right)+\frac{h}{\sqrt{2}}
\left(Y^l-\frac{3C^{l\varphi}v^2}{2\Lambda^2}\right)\right]l_R'+h.c. \,,
\end{align}
where the first and second terms cannot be diagonalized simultaneously.
Rotating the lepton into mass state, 
\begin{equation}
\label{rot_lepton}
l_{L}'=V_{L}l_{L},\quad  l_{R}'=V_{R}l_{R} \,,
\end{equation}
we obtain the Yukawa coupling matrix for $\mathcal{L}_{hll}$ in Eq.(\ref{lageq}) as
\begin{equation}
\label{hll_eff}
y_{ij}=\frac{m_{ij}}{v}\delta_{ij}-\frac{v^2}{\sqrt{2}\Lambda^2}(V_L^\dagger C^{l\varphi}V_R)_{ij},
\end{equation}
which is an arbitrary non-diagonal matrix.
The off-diagonal elements in Eq. \eqref{hll_eff} lead to the hLFV decays $h\to l_i l_j$ 
with branching ratios given by
\begin{equation}
\label{decay_eft}
\text{BR}(h\to l_i l_j)=\frac{m_h}{8\pi \Gamma_h}|\bar{y}_{ij}|^2, \quad \bar{y}_{ij}\equiv \sqrt{|y_{ij}|^2+|y_{ji}|^2}\,,
\end{equation}
where $m_h=125$ GeV is the SM Higgs boson mass, 
and $\Gamma_h\simeq 4.1$ MeV is the total SM Higgs boson decay width. 
When the chiral interactions are written explicitly for BR$(h\to\tau\mu)$, we obtain
\begin{align}
\text{BR}(h\to \tau_L\mu_R)&=\frac{m_h}{8\pi \Gamma_h}|y_{\tau\mu}|^2\,, ~~~~
\text{BR}(h\to \mu_L\tau_R)=\frac{m_h}{8\pi \Gamma_h}|y_{\mu\tau}|^2.
\end{align}

\subsection{Two Higgs Doublet Model}
The 2HDM is a model with an extension of a electroweak scalar doublet. If there is no discrete symmetry that prevents the 
flavor changing neutral current is introduced, then the Yukawa matrices
can no longer be simultaneously diagonalized so flavor-violating interactions are appeared at the tree level. 
Assuming $CP$ conservation,
the 2HDM provides four more additional scalars, i.e., the $CP$-even neutral Higgs boson $H$, 
the $CP$-odd neutral 
Higgs boson $A$ and the charged Higgs bosons $H^\pm$ beside the 
SM-like Higgs boson $h$. 

In the Higgs basis, $H_1$ and $H_2$ can be parameterized as
\begin{equation}
\label{higgsbasis}
H_1=\vecmat{G^+}{\frac{1}{\sqrt{2}}(v+\varphi_1+iG^0)}, \quad H_2=\vecmat{H^+}{\frac{1}{\sqrt{2}}(\varphi_2+iA)}\,,
\end{equation}
where $G^+$ and $G^0$ are the Nambu-Goldstone bosons.

The $CP$-even neutral components $\varphi_1$ and $\varphi_2$ can be rotated to the mass eigenstates $h$ and $H$ by an orthogonal rotation with the mixing angle $\theta_{\beta\alpha}$ as
\begin{eqnarray}
\label{trafo3}
\vecmat{\varphi_1}{\varphi_2}&=&
\begin{pmatrix}
\cos\theta_{\beta\alpha} & \sin\theta_{\beta\alpha}\\
-\sin\theta_{\beta\alpha}& \cos\theta_{\beta\alpha}
\end{pmatrix}
\vecmat{H}{h}\,.
\end{eqnarray}
Note that $h$ is the SM-like Higgs boson for $\sin\theta_{\beta\alpha}\simeq 1$.

The lepton Yukawa sector is given by
\begin{equation}
\label{lag}
-\mathcal{L}_{Y}^{2HDM}
=\barv{L}_L'Y_1l_R'H_1+\barv{L}_L'Y_2l_R'H_2+h.c.\,,
\end{equation}
where $Y_1$ and $Y_2$ are the Yukawa matrices.
The relevant terms to the hLFV decays
after the electroweak symmetry breaking are given as 
\begin{align}
-\mathcal{L}_{Y}^{2HDM}&\supset \barv{l}_L'\frac{vY_1}{\sqrt{2}}l_R'+\barv{l}_L'\left(\frac{Y_1}{\sqrt{2}}\sin\theta_{\beta\alpha}
+\frac{Y_2}{\sqrt{2}}
\cos\theta_{\beta\alpha}
\right)l_R'h
+h.c.
\end{align}
Rotating the leptons into their mass basis, we obtain
\begin{align}
\label{lag2}
-\mathcal{L}_{Y}^{2HDM}&\supset \barv{l}_{L,i} (m_{ij}\delta_{ij})
 l_{R,j} + \barv{l}_{L,i}  \left[\frac{m_{ij}\delta_{ij}}{v}\sin\theta_{\beta\alpha}+\frac{\cos\theta_{\beta\alpha}}{\sqrt{2}}\xi_{ij} \right]l_{R,j}h +h.c.\,,
\end{align}
where 
$m$ is the diagonal lepton mass matrix, 
\begin{equation}
m_{ij}=\frac{v(V_L^{\dagger}Y_1V_R)_{ij}}{\sqrt{2}}=\text{diag}(m_e, m_{\mu}, m_{\tau})_{ij}\,,
\end{equation}
and 
$\xi_{ij} \equiv (V_L^{\dagger}Y_2V_R)_{ij}$, which is generally a nondiagonal matrix.
The second term in Eq. (\ref{lag2}) corresponds to $\mathcal{L}_{hll}$ in Eq.(\ref{lageq}), and
BRs of the hLFV decays are given by
\footnote{The other terms in Eq. \eqref{lag2} induce other LFV Higgs boson decays $H\to l_i l_j$ and $A\to l_il_j$~\cite{Primulando:2016eod, Arganda:2019gnv, Primulando:2019ydt, Hou:2022nyh, Barman:2022iwj}.}  

\begin{equation}
\label{decay_2hdm}
{\rm BR}( h\to l_i l_j)=\frac{m_h \cos^2\theta_{\beta\alpha}}{16\pi\Gamma_h}|\bar{\xi}_{ij}|^2, \quad \bar{\xi}_{ij}=\sqrt{|\xi_{ij}|^2+|\xi_{ji}|^2}.
\end{equation}
The BRs for chiral processes are given by
\begin{align}
\label{chiral_rho}
{\rm BR}( h\to \tau_L\mu_R )&=\frac{m_h \cos^2\theta_{\beta\alpha}}{16\pi\Gamma_h}|\xi_{\tau\mu}|^2\,,~~~~~
{\rm BR}( h\to \mu_L\tau_R)=\frac{m_h \cos^2\theta_{\beta\alpha}}{16\pi\Gamma_h}|\xi_{\mu\tau}|^2.
\end{align}


\section{Analysis at the LHC}
\label{sec:analysis}

\subsection{Polarized taus}
\label{polarization}

At the LHC, constraints on the hLFV couplings $y_{\tau\mu}$ and $y_{\mu\tau}$ are obtained by interpreting the search results of $h\to \tau\mu$ process. In principle, the current studies should already be sensitive to the chirality structure.
To understand that, let us summarize the properties of the $\tau$ decays.
We mainly focus on the case that the $\tau$ leptons decay 
hadronically $\tau \to \,\text{hadrons} + \nu_\tau$ in the $h \to \tau\mu$ events, that is  
\begin{equation}
p p \to h \to \tau^\pm  \mu^\mp\, \to \tau_h^\pm + \nu_\tau + \mu^\mp .
\end{equation}
Here and in the following, we denote the visible components of the hadrons as $\tau_h$, 
which is expected to be identified as a $\tau$-tagged jet at the collider experiment.
In addition to the leptonic decay contributions from the $e\nu$ (17.8~\%) and $\mu \nu$ (17.4~\%) modes, 
main hadronic decay modes of the $\tau$ leptons 
are categorized into the following three groups:
\begin{enumerate}
\item $\pi^\pm$ mode (9.3~\%): $\tau^\pm\to\pi^\pm\nu_\tau$,
\item $\rho^\pm$ mode (25.5~\%):  $\tau^\pm\to\rho^\pm\nu_\tau \to \pi^\pm\pi^0\nu_\tau$,
\item $a_1^\pm$ mode (27~\%): $\tau^\pm \to a_1^\pm\nu_\tau\to \pi^\pm\pi^0\pi^0\nu_\tau$ /$\pi^\pm\pi^\pm\pi^\mp\nu_\tau$.
\end{enumerate}
All three mesons $\pi, \rho$, and $a_1$ are found in a one-prong mode, 
but in principle we can distinguish them by the number of neutral pions, 
meanwhile $a_1$ is also found in a three-prong mode. 
In the all hadronic decay modes, these contributions are 98~\%.
Thus, we consider only these modes.

\begin{figure}[h!]
\includegraphics[width=0.49\textwidth]{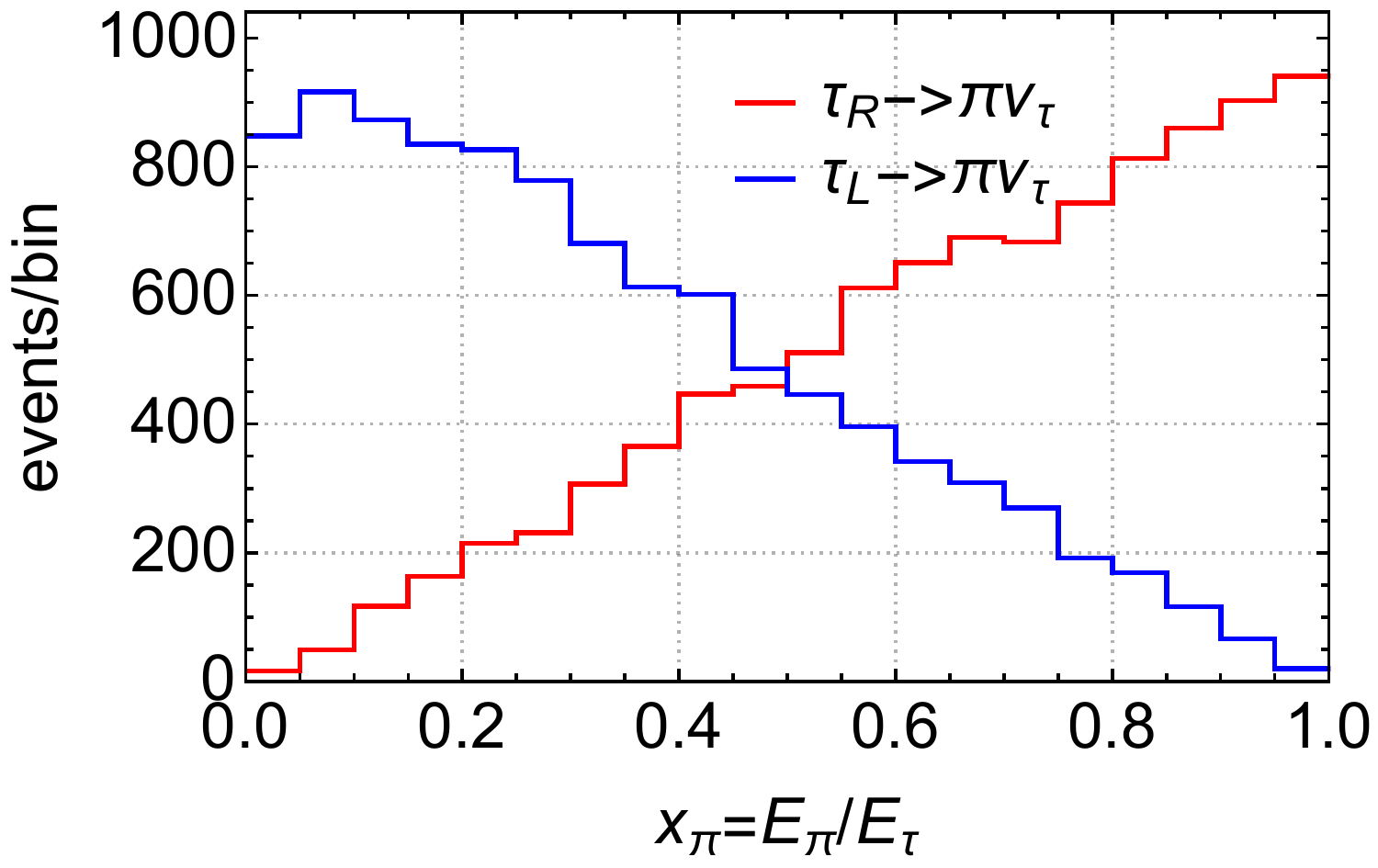}
\includegraphics[width=0.49\textwidth]{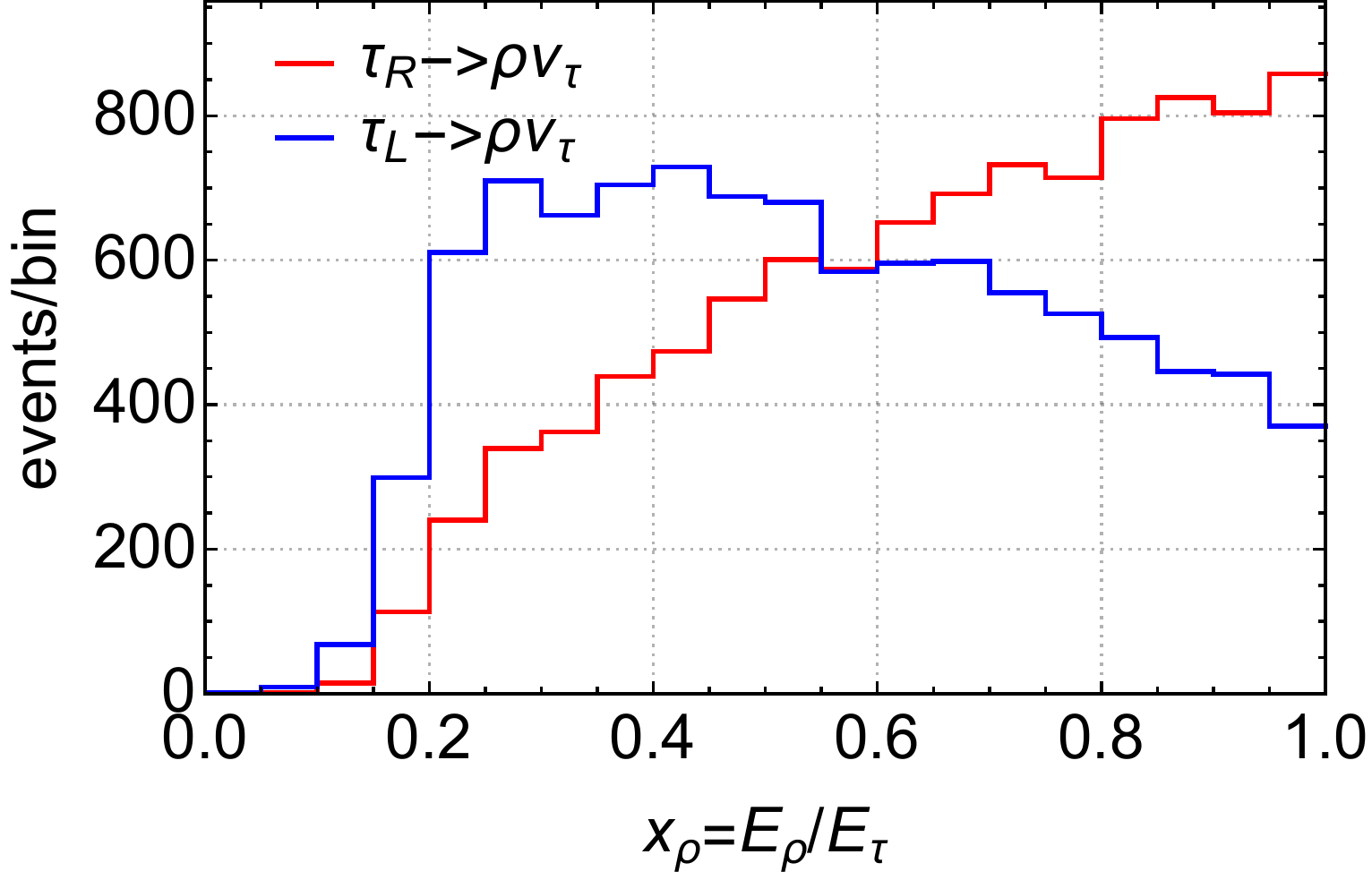}
\includegraphics[width=0.49\textwidth]{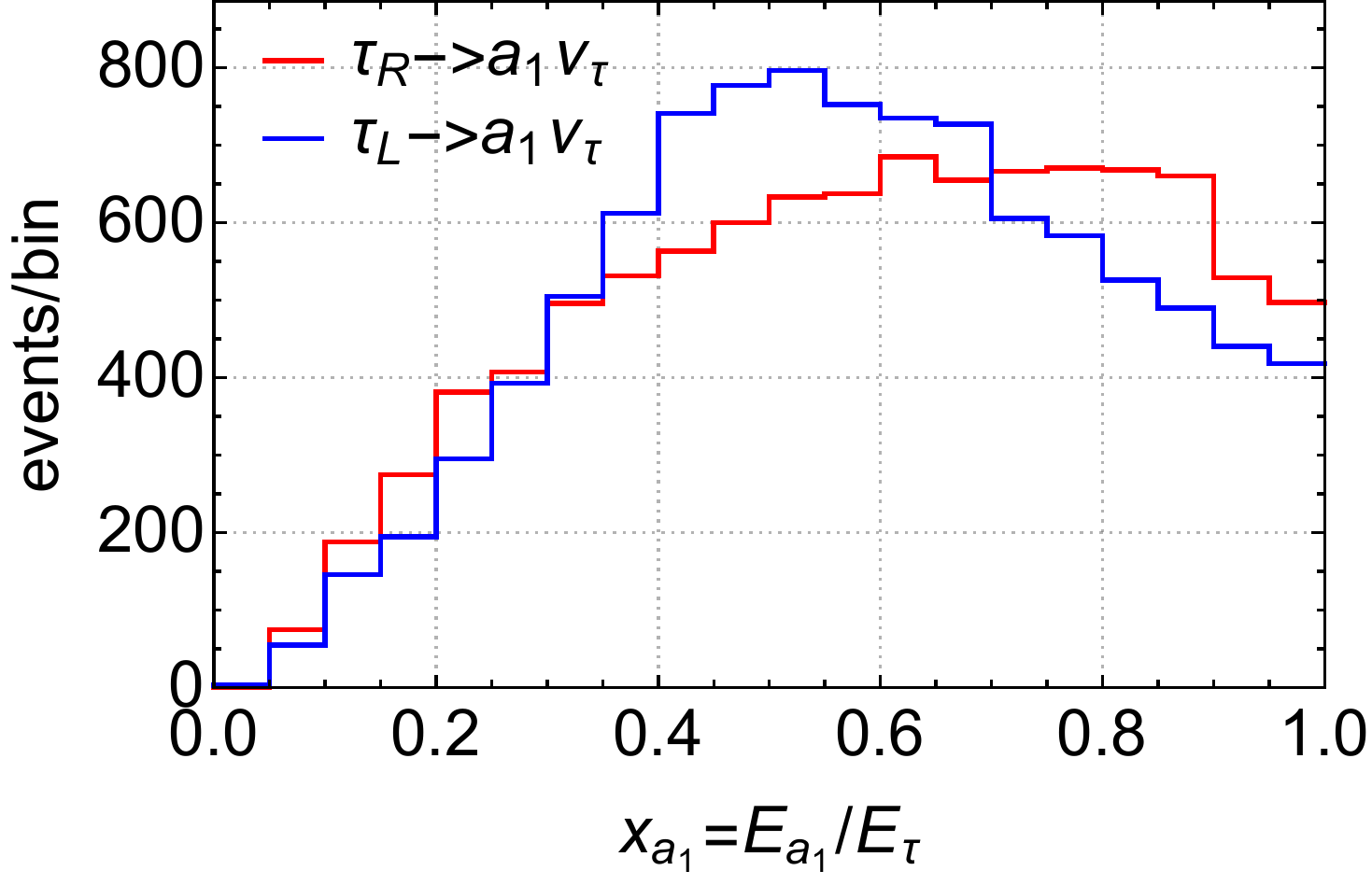}
\includegraphics[width=0.49\textwidth]{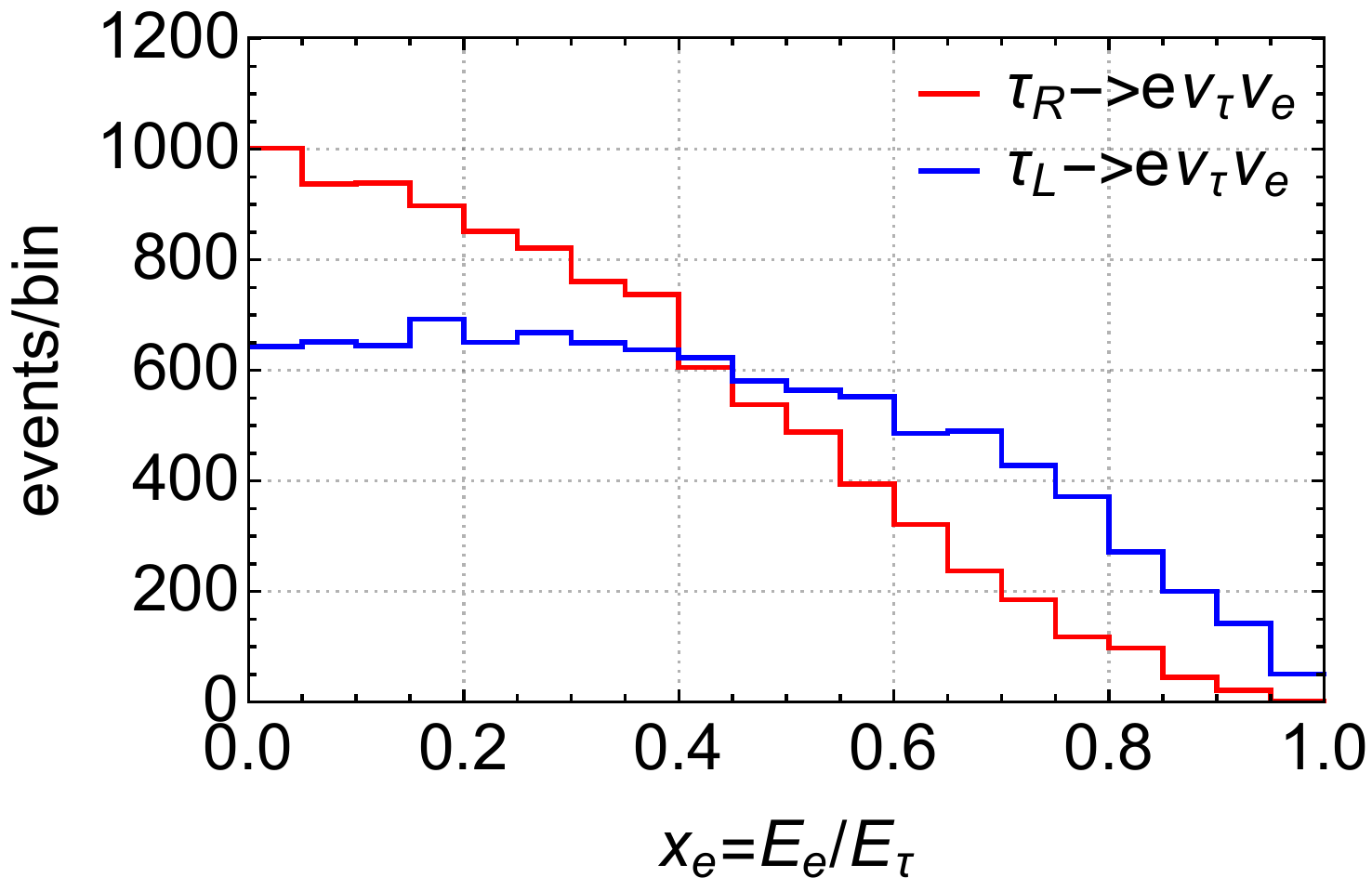}
\caption{The distribution of energy fraction of $\pi,\rho,a_1$ (hadronic) and $e$ (leptonic), i.e. $x_\pi$, $x_\rho$, $x_{a_1}$ as decay product of $\tau$ leptons for different polarizations at the parton level.
}
\label{partondist}
\end{figure}

These hadronic decay modes carry the information on the polarization of $\tau$ leptons~\cite{Bullock:1992yt,Hagiwara:2012vz}, 
i.e. whether $\tau_L$ or $\tau_R$.
One of such simple observables is the fractional energy of $\tau$ decay products $x_i=E_i/E_\tau$ ($i = \pi, \rho, a_1, e$). 
Fig. \ref{partondist} shows the simulated $x_i$ distributions at the parton level for a $\tau_L$ (left-handed $\tau$ lepton),  
and for a $\tau_R$ (right-handed $\tau$ lepton), which are realized by fixing 
$(y_{\mu\tau},y_{\tau\mu})=(0,1)$ and $(y_{\mu\tau},y_{\tau\mu})=(1,0)$ in our effective model, respectively.
The polarized $\tau$ decays is simulated by the package \textsc{TAUDECAY}~\cite{Hagiwara:2012vz}.
In fact these results are consistent with the fractional energy distributions of $\tau$ lepton~\cite{Hagiwara:1989fn, Bullock:1992yt}. 
Among four decay modes shown in Fig. \ref{partondist}, the effect on $\tau$ polarization is most prominent in the $\pi^\pm$ modes.
The $x_\pi$ distribution is hard for $\tau_R$ while it is soft for $\tau_L$. 
For the $\rho$ and $a_1$ modes, $x_\rho$ and $x_{a_1}$ are relatively high for both $\tau_{L/R}$ cases,
and again they are harder for $\tau_R$ than those for $\tau_L$ although they are less sensitive to the polarization.
On the other hand, the $e$-modes provide relatively softer $x_e$ distributions due to the existence of an additional neutrino, 
and harder $x_e$ distribution is obtained for $\tau_L$ than that for $\tau_R$ oppositely to the other hadronic modes.

\begin{figure}[h!]
\includegraphics[width=0.49\textwidth]{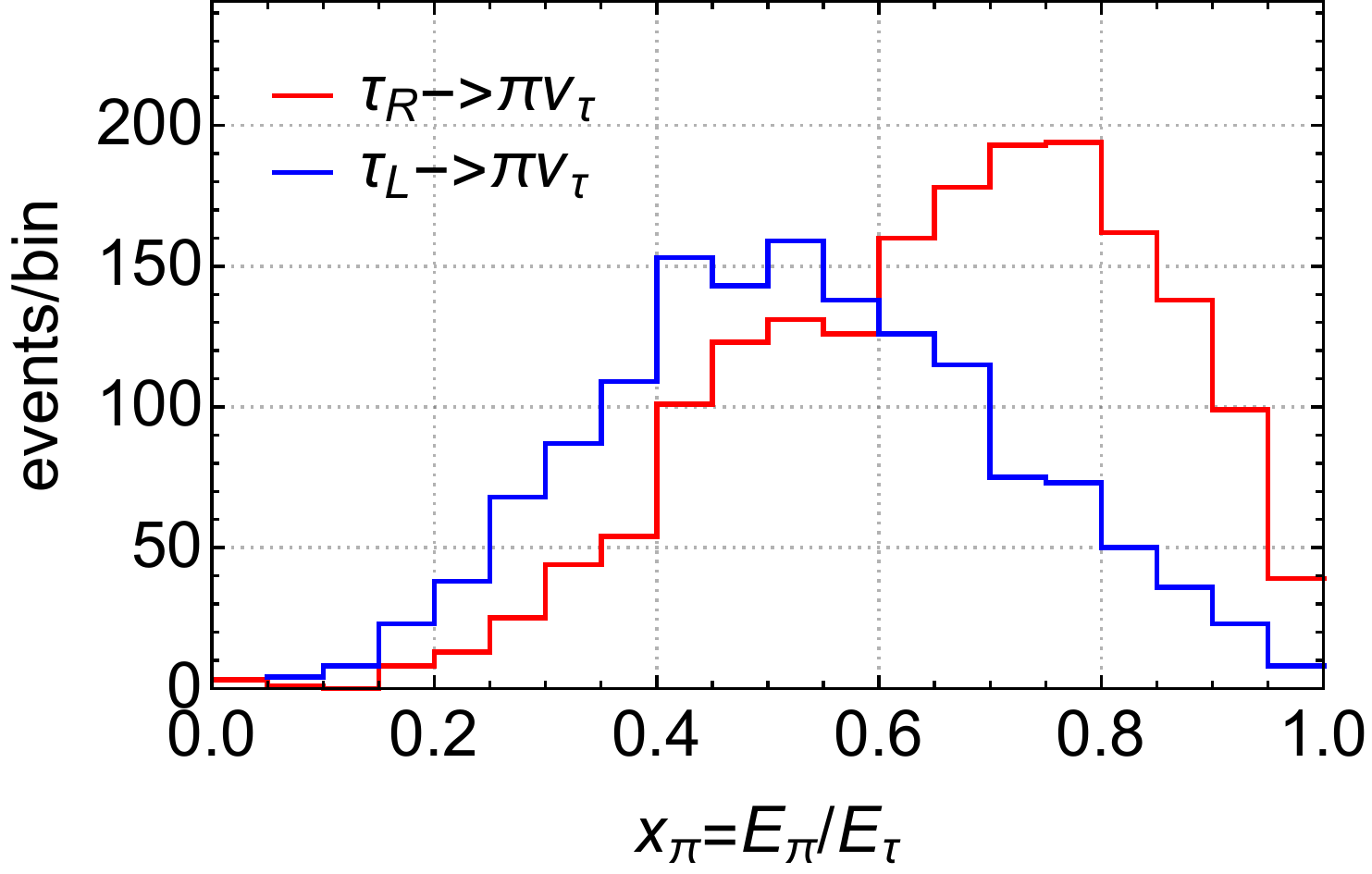}
\includegraphics[width=0.49\textwidth]{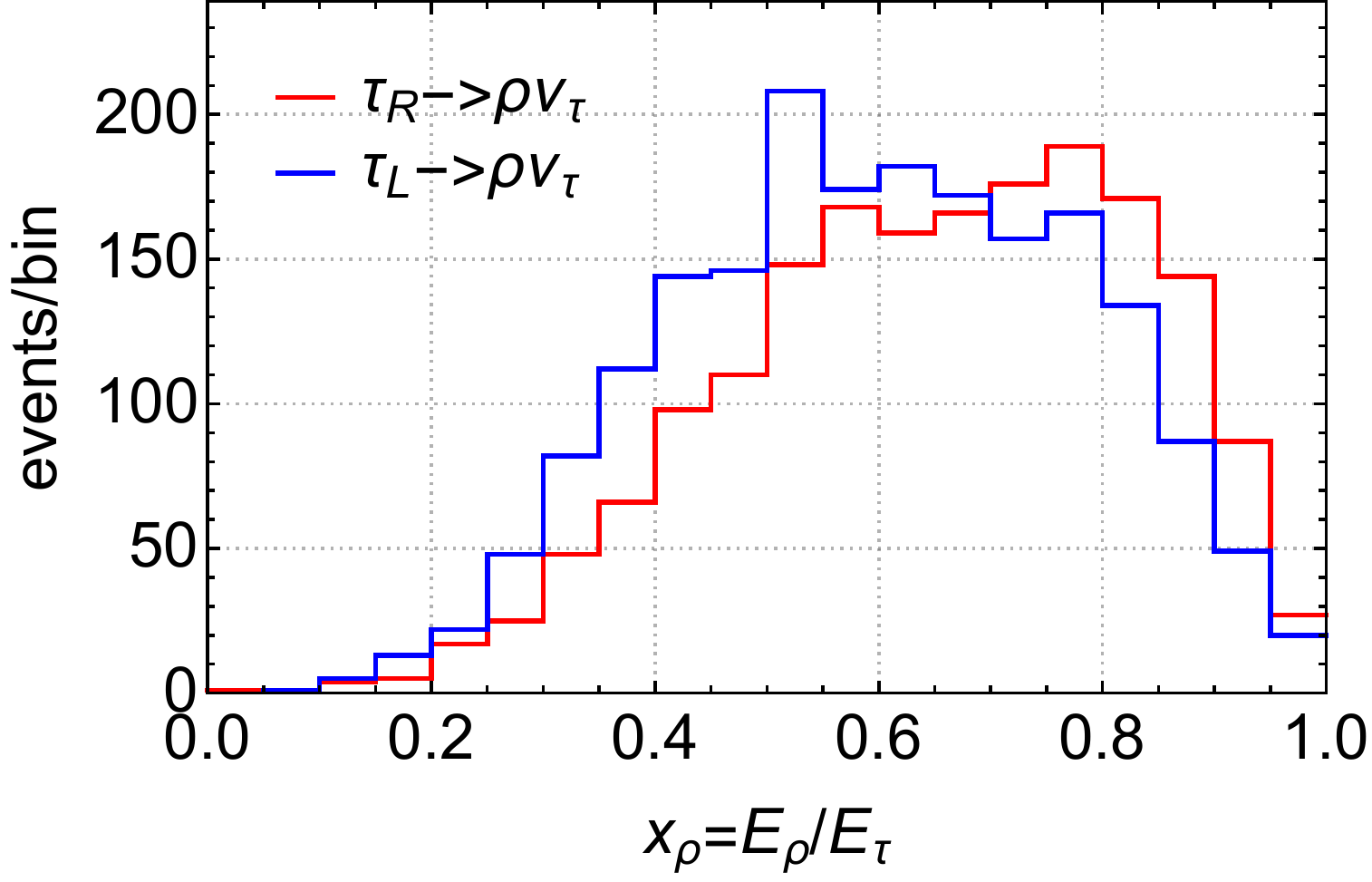}
\includegraphics[width=0.49\textwidth]{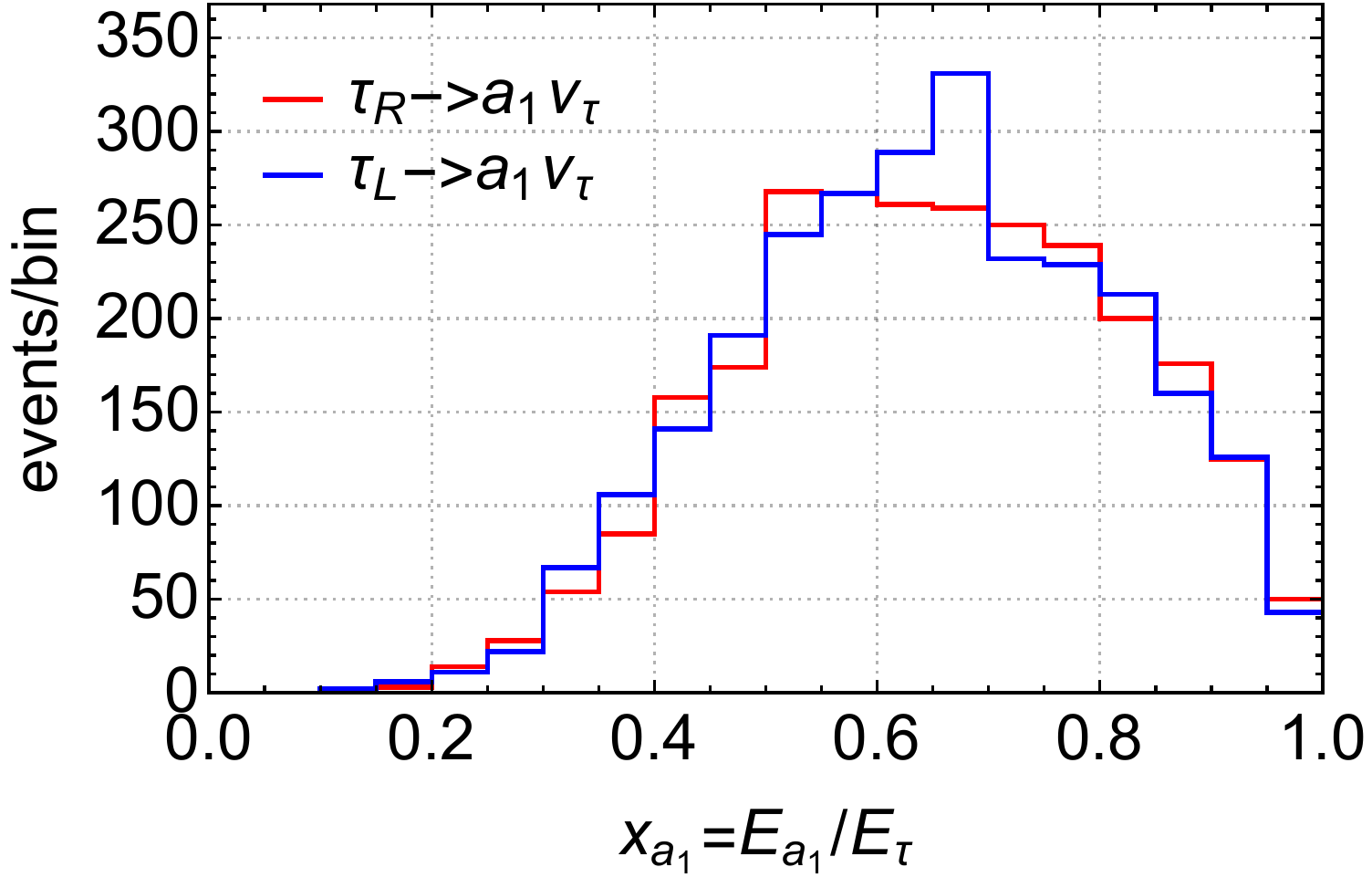}
\includegraphics[width=0.49\textwidth]{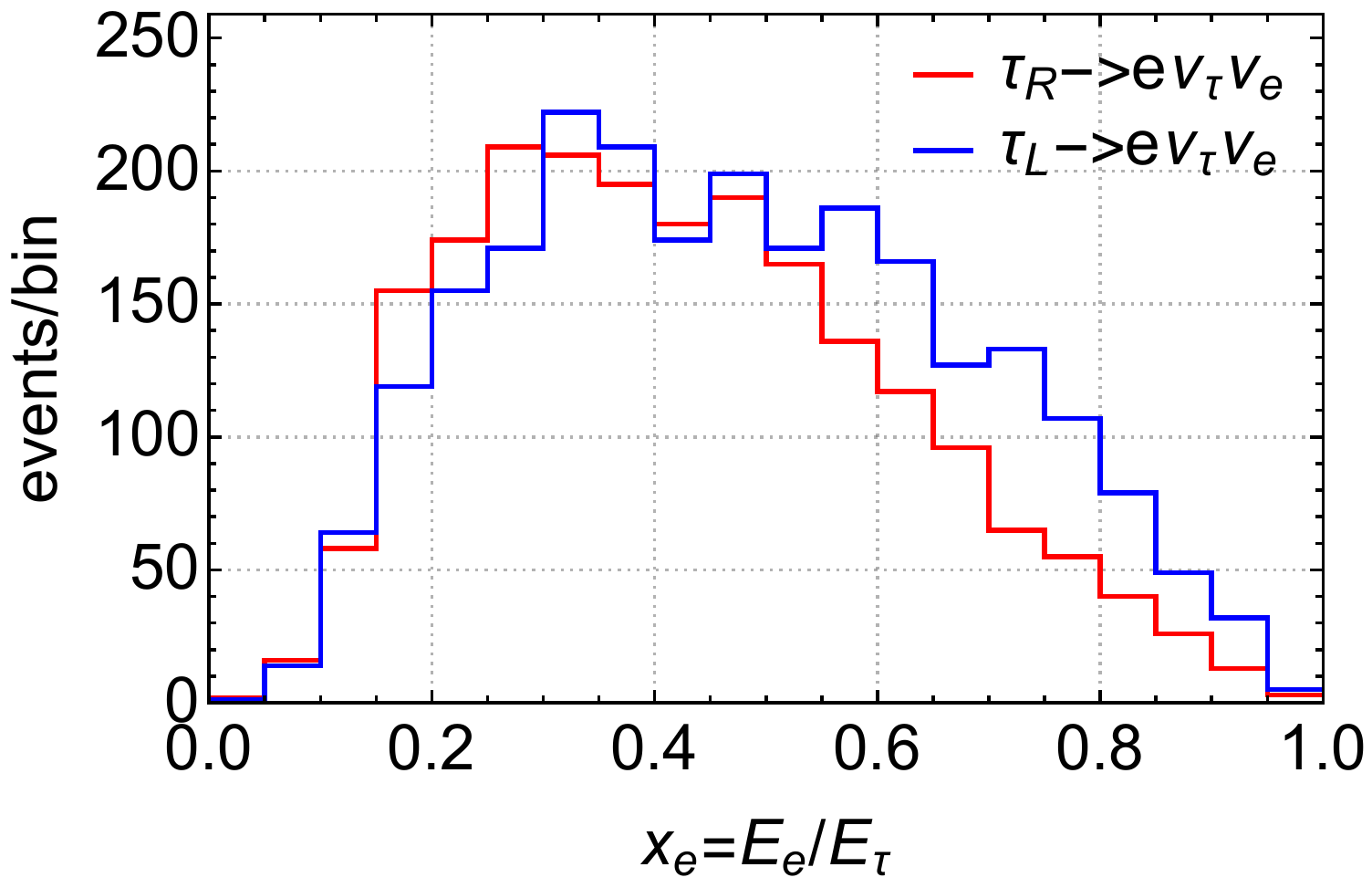}
\caption{The distribution of energy fraction of $\pi,\rho, a_1$ (hadronic) and $e$ (leptonic) as
 decay product of $\tau$ leptons for different polarization at detector level. 
}
\label{recodist}
\end{figure}

Fig. \ref{recodist} shows the corresponding $x_i$ distributions at the reconstructed jet level after the detector simulation.
Although they are smeared, the tendencies discussed above are still visible.
For the jet level observables, we see a significant reduction in lower region of $x_i$ both for $\tau_L$ and $\tau_R$.
This is due to the effects of the jet energy threshold, 
and the events falling down below it cannot be observed.
Therefore, the polarization affects the acceptance of the events 
even though the normalized distributions look similar.

In our analysis, since
${\rm BR}( h \to \mu_L\tau_R) \propto |y_{\mu\tau}|^2$, 
${\rm BR}( h\to \tau_L\mu_R ) \propto |y_{\tau\mu}|^2$
and both can be considered as the independent processes
if we neglect the spin correlation effects, 
any distributions can be expressed as
\begin{equation}
\label{combeq}
f(x;y_{\mu\tau}, y_{\tau\mu})= |y_{\mu\tau}|^2 f_R(x) +  |y_{\tau\mu}|^2 f_L(x).
\end{equation}
Therefore, we discuss the fully polarized cases in this paper: $(y_{\mu\tau}, y_{\tau\mu}) =(1,0)$ and $(0,1)$,
corresponding to the decays into $\tau_R\mu_L$ and $\tau_L\mu_R$, respectively. 
We obtain the results for any general case in the $(y_{\mu\tau}, y_{\tau\mu})$ plane by using the above relationship.

\subsection{Simulation}
\label{sim}

In this section we would like to show how much $\tau$ polarization effects
described above affect the results. Our study is based on the ATLAS study at 36.1~fb$^{-1}$ \cite{ATLAS:2019pmk}.
We also would like to propose an improved analysis strategy 
aiming for identifying the hLFV process $h \to \tau\mu$.
After performing the collider simulation, 
we show that the sensitivity 
in the $(y_{\mu\tau},y_{\tau\mu})$ plane by the  $h\to\tau\mu$ process 
should be asymmetric due to the $\tau$ polarization effects.

Although we would like to directly follow the ATLAS analysis, 
since they provide a results heavily based on the BDT results,
we try to roughly reproduce their results, and discuss the $\tau$ polarization effects based on our simulation. 
They show that at current statistics of 36.1~fb$^{-1}$, the most relevant production modes are still the gluon fusion (ggF) process 
since the vector boson fusion (VBF) modes are statistically still limited. Thus in this paper, we focus on the gluon fusion production mode.
For the final states we focus on the $\tau_h\mu$ channel, since it provides the strongest sensitivity on the hLFV coupling 
among the relevant channels, which include $\mu\tau_e$ and the corresponding VBF channels.
There are many different SM backgrounds for  
the hLFV signal but we only consider the most dominant SM background $Z$ + jets followed by
the decay $Z\to\tau^+\tau^-$, which in the end contribute about a half of the SMBGs 
as shown in ATLAS~\cite{ATLAS:2019pmk} and CMS~\cite{CMS:2021rsq}. 

We generate signal and background events with \textsc{MG5\_aMC@NLO}~\cite{Alwall:2014hca} 
at leading order with $\sqrt{s}$ = 13 TeV. 
The parton shower, hadronization, and the detector response 
are simulated by \textsc{Pythia8}~\cite{Sjostrand:2014zea}
and \textsc{Delphes}~\cite{deFavereau:2013fsa} with the default ATLAS detector card.
The hLFV signal $h\to\tau\mu$ is generated using the 2HDM
where we translated the LFV Yukawa couplings of EFT to the 2HDM term 
via Eq. \eqref{chiral_rho} fixing the  mixing angle $\cos\theta_{\beta\alpha}=0.1$. 
In the following, we show the two cases 
$(y_{\mu\tau},y_{\tau\mu}) \propto (0, 1)$ and  $(1, 0)$ 
for the benchmarks, corresponding to the $\tau_R$ and $\tau_L$ cases, respectively, and mainly show the numbers
for the case of ${\rm BR}( h \to \tau\mu) = 1~\%$. We scale the normalization of the signal samples to reproduce the total ggF Higgs 
production cross section of 48.6~pb, which is the next-to-next-to-next-to-leading order (N$^3$LO) ggF Higgs production cross section at 13~TeV.
Numerically, ignoring the correction to the Higgs total width, 
we adopt the following relation
including the QCD and the other corrections~\cite{LHCHiggsCrossSectionWorkingGroup:2011wcg, Dittmaier:2012vm, LHCHiggsCrossSectionWorkingGroup:2013rie, LHCHiggsCrossSectionWorkingGroup:2016ypw},
\begin{equation}
\label{brsim}
{\rm BR} (h\to\tau\mu)=0.12~\% \times \frac{(|y_{\mu\tau}|^2 + |y_{\tau\mu}|^2)}{10^{-6}}.
\end{equation}
For example, numerically, $\bar{y}_{\mu\tau} = m_\tau/v \simeq 0.0072$ corresponds to ${\rm BR}(h \to \tau\mu) = 6.3~\%$.

For the ggF $\tau_h\mu$ channel, or  non-VBF $\tau_h\mu$ channel, 
we require the following baseline cuts:
exactly one isolated muon and  exactly one $\tau$ jet are required, and their charges are opposite each other.
We rely on {\tt Delphes} for $\tau$ jet identification and we select the working point of the tagging efficiency of 60~\%.
Furthermore, an upper limit on the pseudorapidity difference between $\mu$ and $\tau$ jet, 
$|\Delta\eta(\mu, \tau_{\text{vis}})|<2.0$ is applied to reduce the background from misidentified $\tau$ 
candidates~\cite{ATLAS:2019pmk}. To avoid the missing momentum coming from other sources, 
the sum of the cosine of the angle between $\mu$($\tau_{\rm vis}$) and missing momentum 
in transverse plane is large enough, $\sum_{i=l,\tau_\text{vis}}\cos\Delta\phi(i,\slashed{E}_T)> -0.35$.
The baseline cut is summarized in Table~\ref{baselinecut}.
For the signal samples, we generate 500 000 events per each $\tau_R$, $\tau_L$ sample, while 
1 000 000 events are generated for $Z\to\tau\tau$ background.

\begin{table}[h!]
\begin{tabular}{c|P{8cm}}
\hline
& Selection cuts\\
\hline
\hline
\multirow{4}{*}{Baseline}  & exactly $ 1\mu$ and $1\tau$ jet (opposite sign)\\
&$p_{T,\mu}>27.3$ GeV, \ \ \  $p_{T,\tau_\text{vis}}>25$ GeV\\
&$|\Delta \eta(\mu, \tau_\text{vis})|<2.0$ \\
&$\sum_{i=l,\tau_\text{vis}}\cos\Delta\phi(i,\slashed{E}_T)>-0.35$\\
\hline
\end{tabular}
\caption{Baseline event selection cuts applied for the $\tau_h\mu$ channels.}
\label{baselinecut}
\end{table}

\subsection{Analysis with collinear masses}
\subsubsection{Collinear mass for two missing particles $m_{\text{col2}}$}

The conventional analyses on $h \to \tau \mu$ in literature often follow the analyses 
motivated for the reconstruction of $h\to \tau\tau$ events.
Thus, the most studies rely on the 
reconstructed invariant mass $m_{\tau\tau}$ using the so-called collinear approximation, 
$m_{\text{col}}$~\cite{ATLAS:2019pmk, Elagin:2010aw}
which we explicitly denote $m_{\rm col2}$ in this paper.
The collinear approximation is based on the assumption that 
the momentum of the all invisible decay products of a $\tau$ lepton $\bm{p}_{\tau}^{\text{invis}}$ and 
the momentum of the all visible decay products of a $\tau$ lepton  $\bm{p}_{\tau}^{\text{vis}}$ 
are in parallel to the original $\tau$ momentum, that is, we can express
with the real parameter $x$ as
\begin{equation}
\bm{p}_{\tau}^{\text{vis}} = x \,\bm{p}_{\tau},  \ \ \
\bm{p}_{\tau}^{\text{invis}} = (1-x) \,\bm{p}_{\tau}, \ \ \ {\rm and} \ \ \ 
 \bm{p}_{\tau} = \bm{p}_{\tau}^{\text{vis}} + \bm{p}_{\tau}^{\text{invis}},
\end{equation}
where $x$ describes the fraction of the parent $\tau$’s momenta carried by the visible tau products and $0\le x \le 1$.
This approximation usually works well as long as the original $\tau$ momentum 
$\bm{p}_{\tau}$ is large compared with $m_\tau$.
To reconstruct $m_{\tau\tau}$ in $h\to \tau\tau$ events there is another assumption 
that the missing transverse momentum  $\slashed{\bm{p}}_T$ consists of 
the two neutrinos from the two $\tau$ leptons, which can be written as 
\begin{equation}
\slashed{\bm{p}}_T=c_1 \bm{p}_{T,{\tau_1}}^{\text{vis}}+c_2 \bm{p}_{T,{\tau_2}}^{\text{vis}}\,,
\label{emiss2}
\end{equation}
where $c_1 > 0$ and $c_2 > 0$. We have the relation 
$c_i = (1-x_i)/x_i$ for $i=1,2$.~\footnote{The missing transverse 
momentum is written as $\slashed{\bm{p}}_T$ and 
the transverse missing energy is $\slashed{E}_T= |\slashed{\bm{p}}_T|$.
}
In general, any missing momentum vector can be decomposed into 
the two vector in the transverse plane but not necessarily $c_1>0$ and $c_2 > 0$. 
So we only select such events to perform the collinear approximation.
In this approximation, $\bm{p}_{{\tau_1}}^{\text{vis}}$, $\bm{p}_{{\tau_2}}^{\text{vis}}$ denote 
the two momenta of the visible decay products from the two $\tau$ leptons. 
Then, we can reconstruct the neutrino components by
$\bm{p}_{\nu_i}^{\rm rec}=c_i\bm{p}_{\tau_i}^{\text{vis}}$, 
or the original $\tau$ lepton momentum by 
$\bm{p}^{\rm rec}_{\tau_i}=  \bm{p}_{\tau_i}^{\text{vis}}/x_i$ and the 
invariant mass of the two $\tau$ leptons can be reconstructed as 
\begin{equation}
\label{mcol2}
m_{\text{col2}}^2= (p^{\rm rec}_{\tau_1} + p^{\rm rec}_{\tau_2})^2.
\end{equation}
We can see also the relationship, $m_{\text{col2}}=m_{\text{vis}}/\sqrt{x_1x_2}$ 
where $m_{\text{vis}}^2=(p_{\tau_1}^\text{vis}+p_{\tau_2}^\text{vis})^2$ if we neglect the $\tau$ mass.
This variable $m_{\text{col2}}$ is the $m_{\rm coll}$ variable used in the ATLAS paper~\cite{ATLAS:2019pmk} .
\medskip

\subsubsection{Collinear mass for one missing particle $m_{\text{col1}}$}
Although the above approach is ideal for reconstructing a $\tau\tau$ system, 
such as $h\to \tau\tau$ and $Z \to \tau\tau$ processes,
relying on this variable does not make much sense for reconstructing $\tau \mu$ system
where only one $\tau$ lepton exists.
Thus, we consider another natural variable for detecting $h\to\tau\mu$ process based on the straightforward assumption
instead of Eq.~\eqref{emiss2}, as 
\begin{equation}
\label{emiss1}
\slashed{\bm{p}}_T=c_1 \bm{p}_{T,{\tau}}^{\text{vis}} + c_\perp \bm{\hat{n}}_{T,\perp}, 
\end{equation}
where $\bm{\hat{n}}_{T,\perp}$ denotes a unit vector orthogonal to $\bm{p}_{T,{\tau}}^{\text{vis}}$, 
or $\bm{p}_{T,{\tau}}^{\text{vis}} \cdot \bm{\hat{n}}_{T,\perp}=0$,
in the transverse momentum plane.
Although ideally the second term should vanish in $h\to \tau\mu$ signal events, since there are smearing effects due to, 
for example, the detector response and mismeasurements we introduce the term for 
successfully decompose any missing transverse momentum vector into the two transverse momentum vectors along with $\bm{p}_{\tau}^{\text{vis}}$.

With the parameter $c_1$, we can reconstruct the neutrino momentum and the original $\tau$ lepton momentum as 
\begin{equation}
\bm{p}_{\nu_1}^{\rm rec}=c_1\bm{p}_{\tau_1}^{\text{vis}}, \ \ \ {\rm and}\ \ \ 
\bm{p}_{\tau_1}^{\rm rec}=\bm{p}_{\tau_1}^{\text{vis}}/x_1,
\end{equation}
where we ideally expect $c_1 > 0$, or $0<x_1<1$. Using this momentum we can compute reconstructed $m_{\tau\mu}$ as follows : 
\begin{equation}
\label{mcol1}
(m_{\tau\mu}^{\rm rec})^2 = m_{\text{col1}}^2= (p^{\rm rec}_{\tau_1} + p_{\mu})^2.
\end{equation}
We denote this variable as $m_{\text{col1}}$ and more reasonable for reconstructing a $\tau \mu$ system 
based on the collinear approximation.\footnote{CMS collaboration uses a similar variable $m_{\rm col}$~\cite{CMS:2021rsq}.}
\bigskip

In the following we show the difference between the analysis based on $m_{\rm col1}$ and $m_{\rm col2}$.
We first apply the baseline cut given in the previous section, inspired by the ATLAS analysis, 
and then apply the selection cuts to select only the reasonable events in each 
context of the collinear approximation. 
For the $m_{\rm col2}$ analysis, 
$c_1>0$ and $c_2>0$ should be required for the collinear approximation to be reasonable: however, 
it reduces too many signals. Therefore  we instead require a weaker criteria $x_1 > 0$ and $x_2 > 0$, 
with which $m_{\rm col2}$ is computable. We found this is because the collinear approximation with the two missing particle assumption
is not suitable for reconstructing a $\tau_h \mu$ system, and that is why the sensitivity based on it is worse.
On the other hand, for the $m_{\rm col1}$ analysis, 
we apply $c_1>0$ as the corresponding condition but it is reasonable for reconstructing a $\tau_h \mu$ system.
The number of events for the two signal samples $h\to \tau_R\mu_L$ ($\tau_R$ sample) and $h\to \tau_L\mu_R$ ($\tau_L$ sample),
and $Z\to \tau\tau$ background sample after each step of the selection cuts 
are summarized in Table~\ref{events36ifb}. The numbers are for the integrated luminosity of 36.1~fb$^{-1}$.
We only show the numbers for the cases $\tau_R$ and $\tau_L$ 
but the numbers for the nonpolarized case can be easily obtained 
by taking an average of the numbers in the two columns.

\begin{figure}[h]
\includegraphics[width=0.49\textwidth]{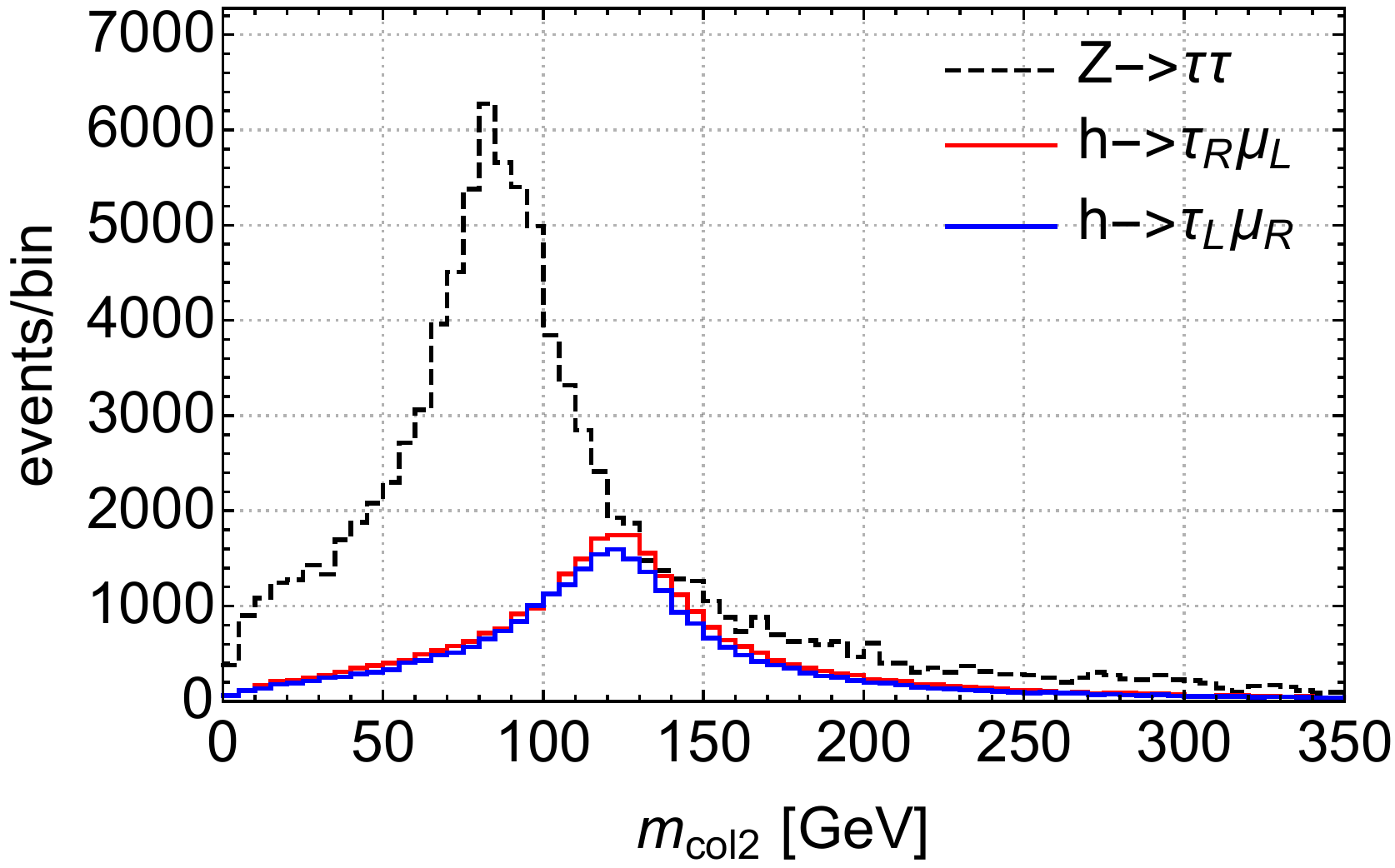}
\includegraphics[width=0.49\textwidth]{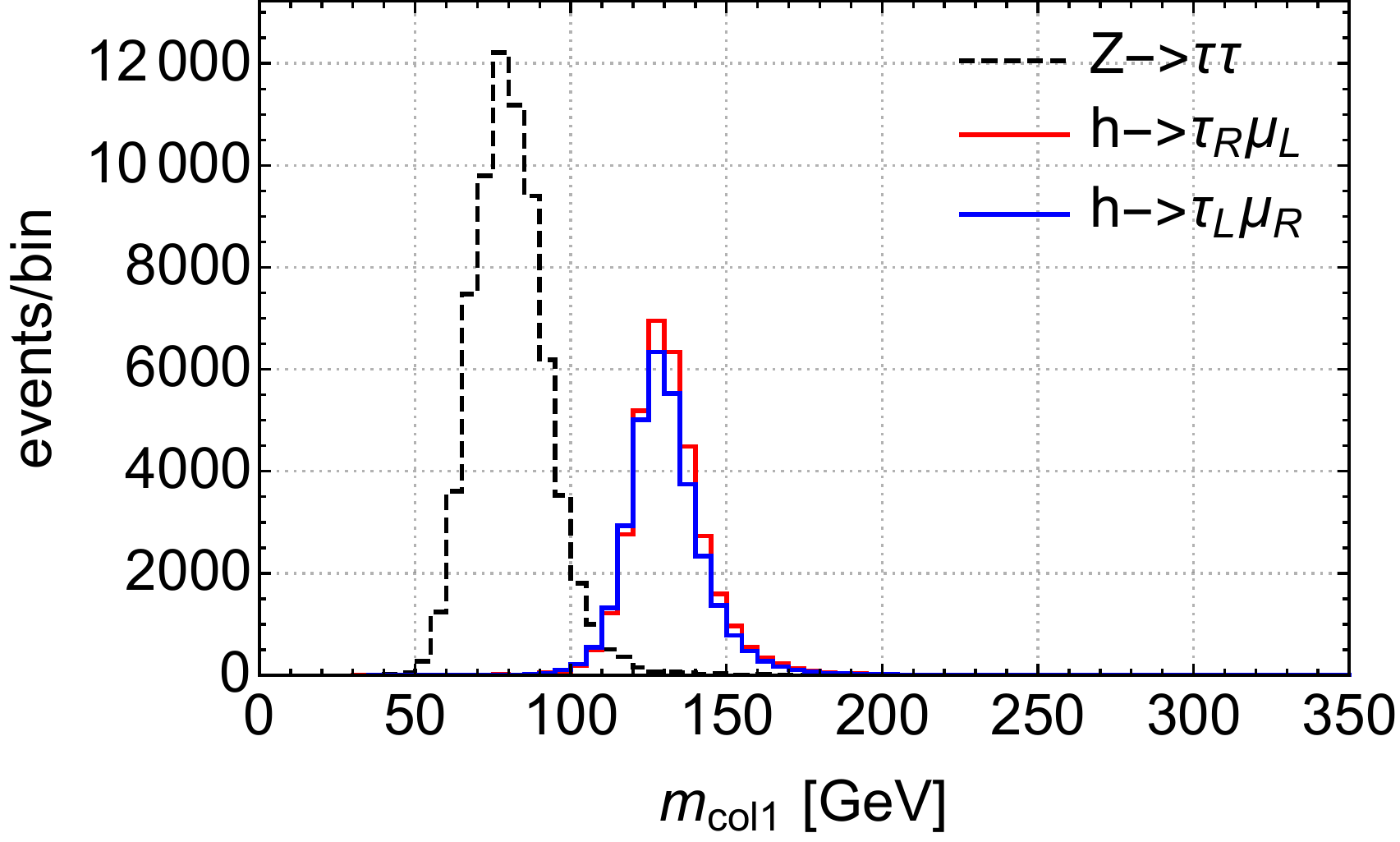}
\caption{
The collinear mass $m_{\text{col2}}$ (left) and $m_{\text{col1}}$ (right) distributions for the signal samples 
$\tau_R$ and $\tau_L$, and $Z\to \tau \tau$ background. 
The signal $m_{\text{col2}}$ ($m_{\text{col1}}$) distributions are scaled with a factor of 20.
}
\label{mcoll2h36dist}
\end{figure}

The $m_{\text{col}2}$ and $m_{\text{col}1}$ distributions at 36.1~fb$^{-1}$ after the appropriate selection cuts 
are given in Fig.~\ref{mcoll2h36dist} in the left and right panels, respectively. 
The distributions for the signal samples $\tau_R$ (red), $\tau_L$ (blue) 
and the $Z\to \tau\tau$ sample (black dotted) are shown.
We see that  both variables $m_{\text{col}2}$ and  $m_{\text{col}1}$ have a peak at $m_h=125$~GeV.
The $m_{\text{col}1}$ peak is very sharp not only for the signal but also for the $Z\to \tau\tau$ BG 
and we see a clear separation between the signal  against the background distributions.
On the other hand, $m_{\text{col}2}$ distributions provide a broader peak and 
exhibit a significant overlap between the signal and background.
Thus, we can reduce the background events by selecting the peak region keeping the signal evens, 
and more efficiently by $m_{\rm col1}$ than by $m_{col2}$.

We define the signal regions as $|\Delta m_{\text{col}i}|= |m_{\text{col}i} - m_h| \le \Delta m_{\text{col}i}^{\rm th}$, 
and show the results for the three possible choices of 
$\Delta m_{\text{col}i}^{\rm th} = 25$, 10, and 5~GeV. 
The numbers after selecting those signal regions are summarized in Table~\ref{events36ifb}. 
We observe that by selecting $m_{\rm col2}$ region with 25~GeV width,
the background can be reduced up to 0.23~\% while keeping the signal only about $5-6$~\%.
On the other hand, by selecting $m_{\rm col1}$ region with 25~GeV width,
the background can be reduced down to 0.04~\% while keeping the signal about $12-13$~\%.
Therefore, using $m_{col1}$ variable would provide a significant improved signal to background ratio.
Another important observation is $m_{\rm col1}$ distribution provides a sharper peak, so in principle 
selecting a narrower signal region would improve the signal over background ratio, which one can explicitly 
see from Table~\ref{events36ifb} and only seen in $m_{\rm col1}$ analysis.

\begin{table}[t]
\centering
\begin{tabular}{ll|cc|c||c|cc}
\hline
&& \multicolumn{2}{c|}{$h\to \tau\mu_{(BR=1~\%)}$ } &&& \multicolumn{2}{c}{BR$^{95\%}$} \\
&& $\tau_R $ & $\tau_L$ & $Z \to \tau_h \tau_\mu$ & $N^{95\%}$   & ${\tau_R}$ & ${\tau_L}$ \\ 
\hline
&$\sigma$ at 13 TeV LHC & \multicolumn{2}{c|}{$355$~fb} & 258~pb && \\
&for ${\cal L} =36.1$fb$^{-1}$  & \multicolumn{2}{c|}{12795} &$ 9.31 \times 10^6$ &&\\
\hline
&baseline cuts & 1979 & 1742 & 130147 &\\
\hline
\hline		

\multirow{4}{*}{$m_{\text{coll}2}$}
& $x_1>0$ and $x_2>0$ & 1672 & 1480 & 102536 & 747 & 0.45  & 0.50 \\
\cline{2-8}
&$|m_{\text{coll}2}  - m_h  |< 25$~GeV  &  717&  643 & 21473 & 342 &  0.48 & 0.53\\
&$|m_{\text{coll}2}  - m_h  |< 10$~GeV& 344 & 304 & 7639 & 204  & 0.59 & 0.67 \\ 
&$|m_{\text{coll}2}  - m_h  |<  5$~GeV& 177 & 157 & 3776 & 143  & 0.81 & 0.91 \\ 
\hline
\hline
\multirow{4}{*}{$m_{\text{coll}1}$}
&$c_1>0$ & 1765 & 1608 & 68602 & 610 & 0.34 & 0.38 \\
\cline{2-8}
&$|m_{\text{coll}1}  - m_h |< 25$~GeV &  1626 &  1493 & 4023 & 148 & 0.091& 0.099 \\
&$|m_{\text{coll}1}  - m_h  |< 10$~GeV & 1080 & 1008 & 639 & 58.9 & 0.055 & 0.059 \\
&$|m_{\text{coll}1}  - m_h  |< 5$~GeV &617 & 577 & 216 & 34.2 & 0.056 & 0.059\\
\hline
\end{tabular}
\caption{Cut flow for $m_{\rm coll2}$ and $m_{\rm coll1}$ analyses. The number of the signal $h \to \tau_h\mu$ $(\tau_R, \tau_L)$ and $Z$ background are shown for ${\cal L} = 36.1$~fb$^{-1}$. Estimated upper bound for each signal region, $N^{\rm 95\%}= 1.65 \sqrt{N_b}$ 
($N_{b} = 2 N_{Z\to\tau\tau}$), and the corresponding sensitivity values for ${\rm BR}( h \to \tau \mu)$.
}
\label{events36ifb}
\end{table}

Note that our baseline selection cuts are inspired by the cuts given by the ATLAS analysis,
and for the $m_{\rm col2}$ analysis the signal and $Z\to\tau\tau$ background 
numbers after requiring $x_1>0$ and $x_2>0$ are slightly larger 
but rather consistent with the numbers given in Table~5 in the ATLAS paper. 
We assume the total background contributions as the twice of the $Z\to \tau\tau$ background contributions 
following the same table, $N_b = 2 N_{Z\to \tau\tau}$.
We estimate the 
95\%~C.L. upper bound of the signal number in a certain signal region as 
$N^{\rm 95\%}= 1.65 \sqrt{N_b}$, where we employ a frequentist approach to obtain the one-side 95\%~C.L. interval.
Then,  from $N^{\rm 95\%}$  in each signal region, we estimate $BR^{95\%}$, 
the 95\%~C.L. upper bound for the branching ratio ${\rm BR}( h \to \tau\mu)$, based on the expected signal numbers. 
Since the expected signal numbers depends on the signal assumptions $\tau_R$ or $\tau_L$, or the $\tau$ polarization,
the corresponding $BR^{95\%}$ also depends on it.
Note that we do not consider the systematic uncertainty to estimate the sensitivity, thus, a large signal over background ratio 
is important for those numbers to be reliable since the uncertainty effects become relatively small.
Once one obtains the $BR^{95\%}$, the corresponding $\bar{y}_{\tau\mu}^{95\%}$ can be calculated straightforwardly using Eq.~\eqref{brsim}. 
\medskip

Next, let us discuss the $\tau$ polarization effects. 
Already at the step at the baseline cuts, 
the signal efficiencies for $\tau_R$ and $\tau_L$ are different, and the $\tau_R$ sample survives more efficiently.
The difference is about $\pm$ 6\% from the nonpolarized case, which is understood by the effects of the jet $p_T$ threshold.
For the $m_{\rm col2}$ analysis, the difference is kept about $\pm 6\%$ after selecting the events with 
$x_1 > 0$ and $x_2 > 0$, and also the case after further selecting the $m_{\rm col2}$ peak region.
Thus, interpreting the results based on $m_{\rm col2}$ analysis is sensitive to the polarization about 6~\%.
On the other hand, for $m_{\rm col1}$ analysis the difference becomes diminished to $\pm 4.6\%$, and 
further weakened around the $m_{\rm col1}$ peak region to $\pm 3.3\%$. 
Final sensitivity is less sensitive when we use the $m_{\rm col1}$ but asymmetric behavior still exists.

\begin{figure}[h!]
\centering
\includegraphics[width=0.75\textwidth]{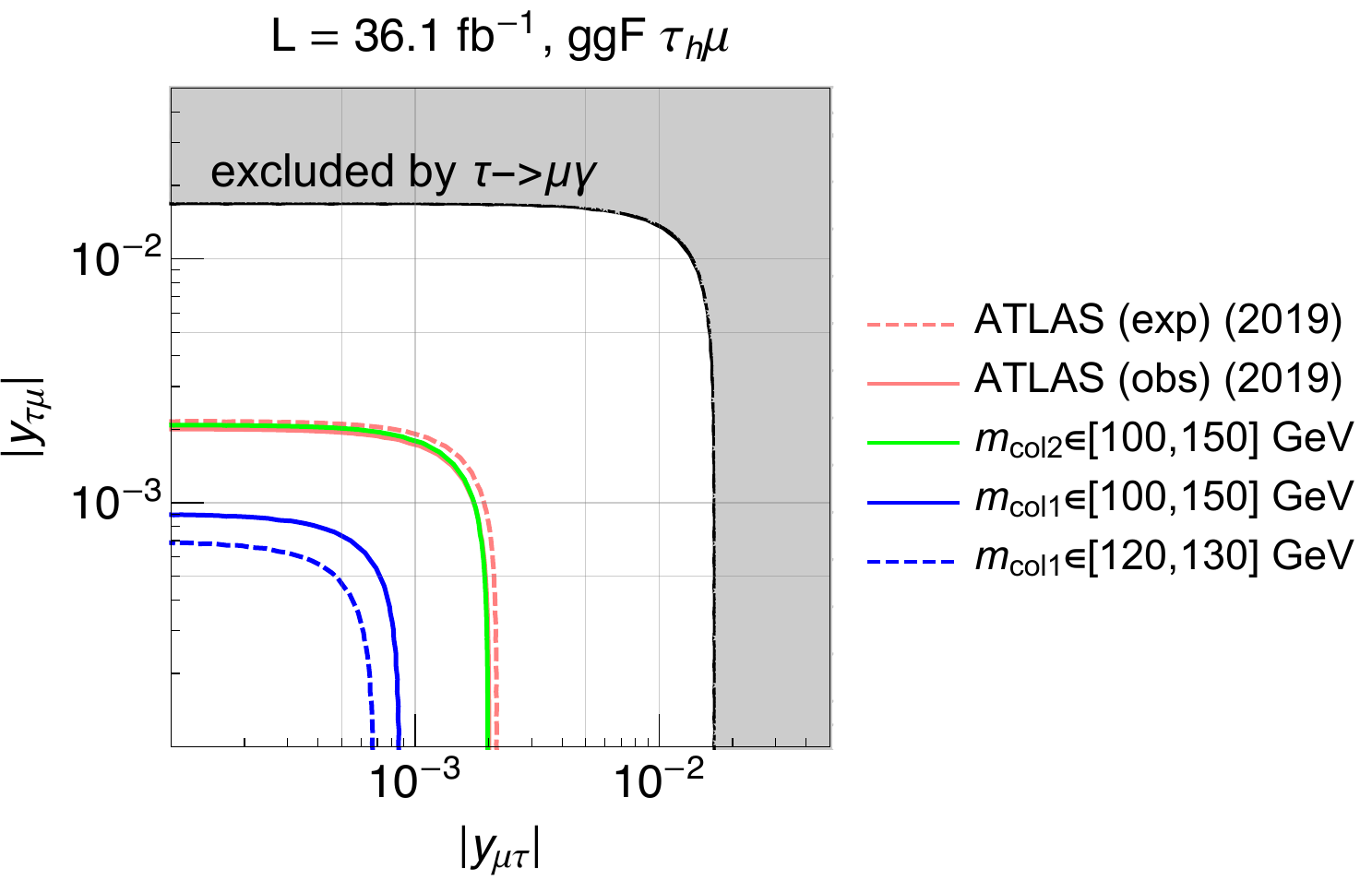}
\caption{Estimated upper bound on the flavor-violating Yukawa couplings $|y_{\tau\mu}|$,$|y_{\mu\tau}|$ 
using $h\to\tau_h \mu$ signals based on $m_{\rm col2}$ and $m_{\rm col1}$ analyses for integrated luminosity 
${\cal L}=36.1$~fb$^{-1}$. 
The ATLAS results are also shown for reference.
The indirect limits from $\tau \to \mu\gamma$ searches~\cite{Harnik:2012pb,BaBar:2009hkt} are indicated as the shaded region.
}
\label{2dymcoll2}
\end{figure}

Fig. \ref{2dymcoll2} shows
our sensitivity results  for the integrated luminosity of 36.1~fb$^{-1}$ expressed in the $y_{\mu\tau}$-$y_{\tau\mu}$ plane, 
based on $m_{\rm col2}$ and $m_{\rm col1}$ analyses using only $\tau_h \mu$ modes.
Unless the spin correlation effects are important, the contour of the exclusion boundary becomes an ellipse
interpolating the values estimated by the two extreme cases $\tau_R$ and $\tau_L$.

The green line shows the sensitivity based on the $m_{\rm col2}$ analysis, using the signal region of $\Delta m_{\rm col2}^{\rm th}= 25$~GeV, 
while the blue solid (dashed) line shows that based on the $m_{\rm col1}$ analysis for the $\Delta m_{\rm col1}^{\rm th} = 25$~GeV  (5~GeV). 
By changing the analysis from $m_{\rm col2}$ to $m_{\rm col1}$, 
the sensitivity is significantly improved by a factor of 5 in terms of the constraints on the branching ratio.
If we consider an even narrower signal region with $\Delta m_{\rm col1}^{\rm th} = 5$ and $10$~GeV, 
the sensitivity is further improved by a factor of 10 compared with the $m_{\rm col2}$ analysis.
It is thanks to the distinct peak shape of the $m_{\rm col1}$ distribution for the signal.
Although we need to consider seriously whether the experimental smearing effects spoil this property or not, 
using $m_{\rm col1}$ would be advantageous since $m_{\rm col2}$ distribution does not have such a property.
One can see that taking a narrower signal region does not improve the sensitivity for $m_{\rm col2}$ analysis.

Our estimate of the sensitivity by the $m_{\rm col2}$ analysis is close to the expected sensitivity in
the non-VBF $\tau_h\mu$ mode of BR($h\to\tau\mu$) = 0.57~\%~$(\bar{y}_{\tau\mu}=0.0022)$,
which is shown in the red dashed line.
Although we expect our sensitivity by a simple cut based analysis is worse than their sophisticated BDT analysis, 
these numbers coincide accidentally, which would be understood because we do not take the uncertainty into account.
However, since the assumption of the setting for $m_{\rm col2}$ analysis and 
that for $m_{\rm col1}$ analysis are the same within our analysis, the improvement by using the 
$m_{\rm col1}$ for the analysis should persist.
\medskip

The resulting sensitivity contours become not circles but ellipses  in $(y_{\tau\mu},y_{\mu\tau})$ plane 
when one includes the polarization effects appropriately. 

Essentially their results are only rigorously correct along the line $y_{\tau\mu}=y_{\mu\tau}$ 
and the polarization effects would modify $\pm$~$4-6$~\% in the branching ratio, 
and $\pm$~$2-3$~\% in $(y_{\mu\tau},y_{\tau\mu})$ plane.
The current ATLAS bound based on the non-VBF $\tau_h\mu$ mode would also be modified.

According to the ATLAS analysis, combining all other modes of $h\to\tau\mu$ process improves the sensitivity about 35~\%, 
which gives the sensitivity down to BR($h\to\tau\mu$)= 0.37~\%~$(\bar{y}_{\tau\mu}=0.0017)$. 
The improvement factor of $5-10$ is obtained only by changing the analysis strategy not by considering the other modes.
Combining the sensitivity for the other modes applying $m_{\rm col1}$ analysis would improve the sensitivity further, 
and we expect it also 35~\% as a reference value,
although we leave it for a future work since we need to perform a further study to confirm it.

\subsection{Future prospect}
\subsubsection{Ultimate sensitivity}

The sensitivity would scale with the integrated luminosity as proportional to $1/\sqrt{\cal L}$, 
since we employ the formula $N^{95\%} = 1.65 \sqrt{N_b}$ for estimating it.
The estimated upper limits on the branching ratio and the corresponding $\bar{y}_{\tau\mu}$ values 
for the several assumptions on the integrated luminosity $\cal L$ are summarized in Table~\ref{citable}.
They are estimated by using the $\tau_h\mu$ channel only, 
and we would expect 35~\% improvements by combining all the other modes.
As explained before the $\tau_R$ sample gives a more stringent bound than the $\tau_L$ sample,
and the effects are about $\pm~4~\%$  in the branching ratio, 
and $\pm ~2~\%$ in the $\bar{y}_{\mu\tau}$ value.

\begin{figure}[h]
\centering
\includegraphics[width=0.65\textwidth]{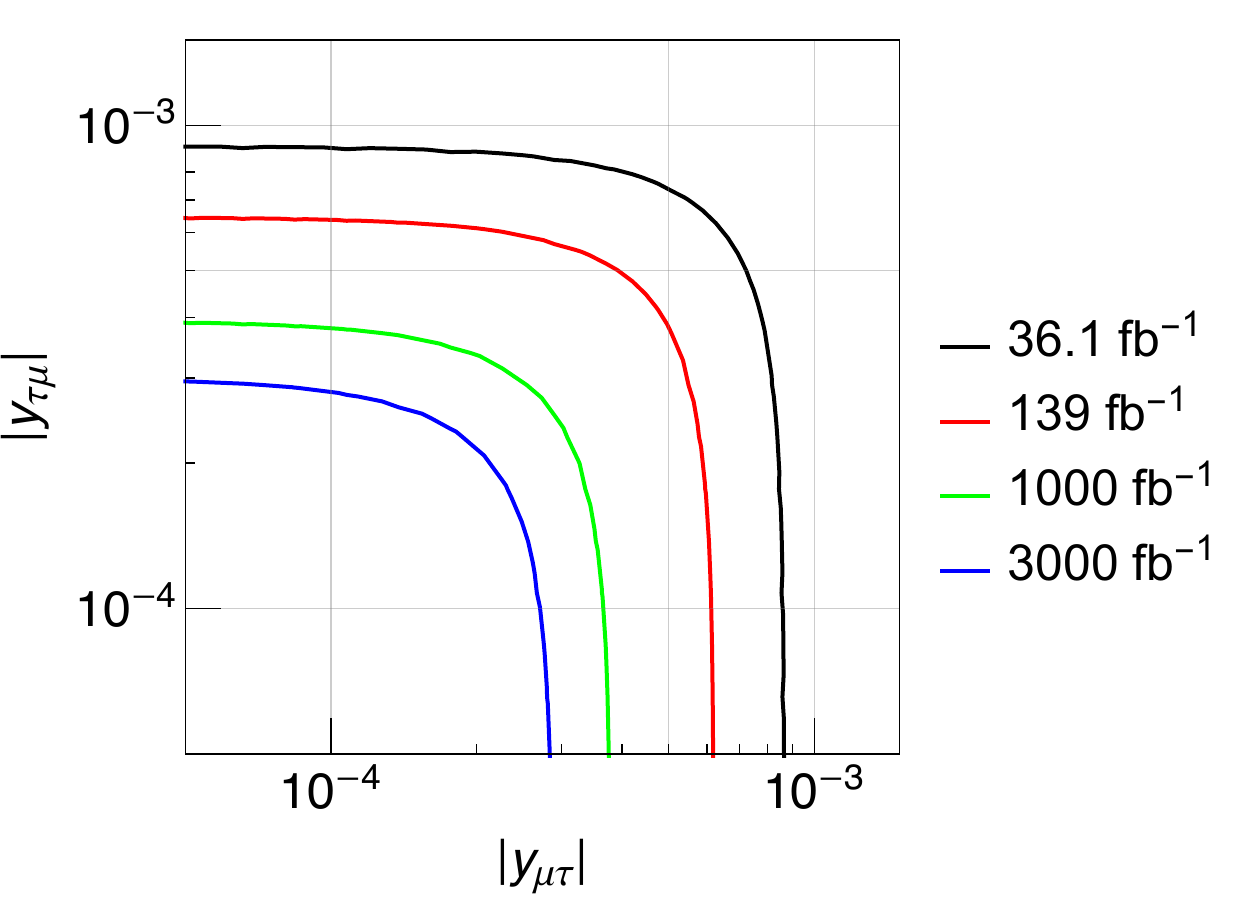}
\caption{Future prospects of the estimated sensitivity for the Yukawa couplings in $|y_{\mu\tau}|$,$|y_{\tau\mu}|$ plane 
based on the $m_{\rm col1}$ analysis with $\Delta m_{\rm col1}^{\rm th} = 25$~GeV using $h \to \tau_h \mu$ process.
The expected results for the integrated luminosity of 36.1 fb$^{-1}$, 139~fb$^{-1}$, 1~ab$^{-1}$, and 3~ab$^{-1}$ are shown.
}
\label{fig:future}
\end{figure}

This information is depicted in Fig.~\ref{fig:future}. 
We estimate that 
the upper bound $\bar{y}_{\tau\mu}^{95\%}$  at HL-LHC can be up to $3 \times 10^{-4}$, corresponding 
to BR($h\to\tau\mu$)$\sim 10^{-4}$.
Our result is in the same order of the  result obtained in Ref.~\cite{Barman:2022iwj}.

\begin{table}[h!]
\centering
\begin{tabular}{lc|l|llc|ccc}
\hline
$\Delta m_{\rm col1}^{\rm th}$ & ${\cal L}$ [fb$^{-1}$] & $N^{95\%}$ & \multicolumn{3}{c|}{BR$^{95\%}$ [$\times 10^{-4}$]}
& \multicolumn{3}{c}{$\bar{y}_{\tau\mu}^{95\%} [\times 10^{-4}$]}\\
&&& $\tau_R$ & $\tau_0$ & $\tau_L$ & $y_{\mu\tau}$ & $y_{\mu\tau}=y_{\tau\mu}$ & $y_{\tau\mu}$\\
\hline
\hline
\multirow{4}{*}{25~GeV} & 36.1 & 148 &9.10  & 9.51  & 9.91  & 8.68  & 8.88  & 9.06 \\
&139 & 290 &4.64  & 4.85 & 5.05  & 6.20  & 6.34  & 6.47 \\
&1000  &779 &1.73  & 1.81  & 1.88  & 3.79  & 3.87  & 3.95 \\
&3000  & 1349 &0.99  & 1.04  & 1.09  & 2.88  & 2.94  & 3.00  \\
\hline
\multirow{4}{*}{5~GeV} & 36.1 & 34.2 &5.55  & 5.74  & 5.93  & 6.78  & 6.90  & 7.01 \\
&139  & 67.1 &2.83  & 2.93  & 3.02  & 4.84  & 4.92  & 5.01 \\
&1000& 180 &1.06  & 1.09  & 1.13  & 2.96  & 3.01  & 3.06\\ 
&3000   & 312 &0.61  & 0.63  & 0.65  & 2.25  & 2.28  & 2.32 \\
\hline

\end{tabular}
\caption{Estimated sensitivity for the several assumed integrated luminosities ${\cal L}= 36.1, 139, 1000$, and $3000$~fb$^{-1}$.
The $N^{95\%}$ in the signal region, the corresponding sensitivities on the BR and $\bar{y}_{\mu\tau}$ are given.
The two cases for $\Delta m_{\rm col1}^{\rm th} = 25$ and  $5$~GeV are shown.
The unpolarized case $y_{\tau\mu} =y_{\mu\tau}$ is denoted as $\tau_0$.
}
\label{citable}
\end{table}

\subsubsection{Sensitivity for the chirality structure}

We would like to show how much sensitive to the chirality structure when we find the finite number of signals.
For that we would like to show we can use the $x_1$ distribution, which naturally obtained along with computing the $m_{\rm col1}$ variable.
To illustrate the procedure, let us take BR$(h\to\tau\mu)=0.12~\%$, corresponding to $\bar{y}_{\mu\tau} = 10^{-3}$ as a benchmark scenario,
which is close to the best fit value for the recently reported excess by the ATLAS analysis with 138~fb$^{-1}$~\cite{ATLAS:2022conf}.
We consider the three cases keeping $\bar{y}_{\mu\tau} = 10^{-3}$: 
the purely $\tau_R$ case, the purely $\tau_L$ case, and the nonpolarized ($\tau_0$) case,
and for each case we assume that the expected number of events are exactly observed: 
\begin{align}
(\hat{y}_{\mu\tau}, \hat{y}_{\tau\mu}) = \begin{cases}
(10^{-3}, 0)  &[\tau_R\ {\rm scenario}],\\
(0,10^{-3})   &[\tau_L\ {\rm scenario}],\\
(7.1\times 10^{-4}, 7.1\times 10^{-4}) &[\tau_0\ {\rm scenario}],
\end{cases}
\end{align}

\begin{figure}[h]
\centering
\includegraphics[width=0.49\textwidth]{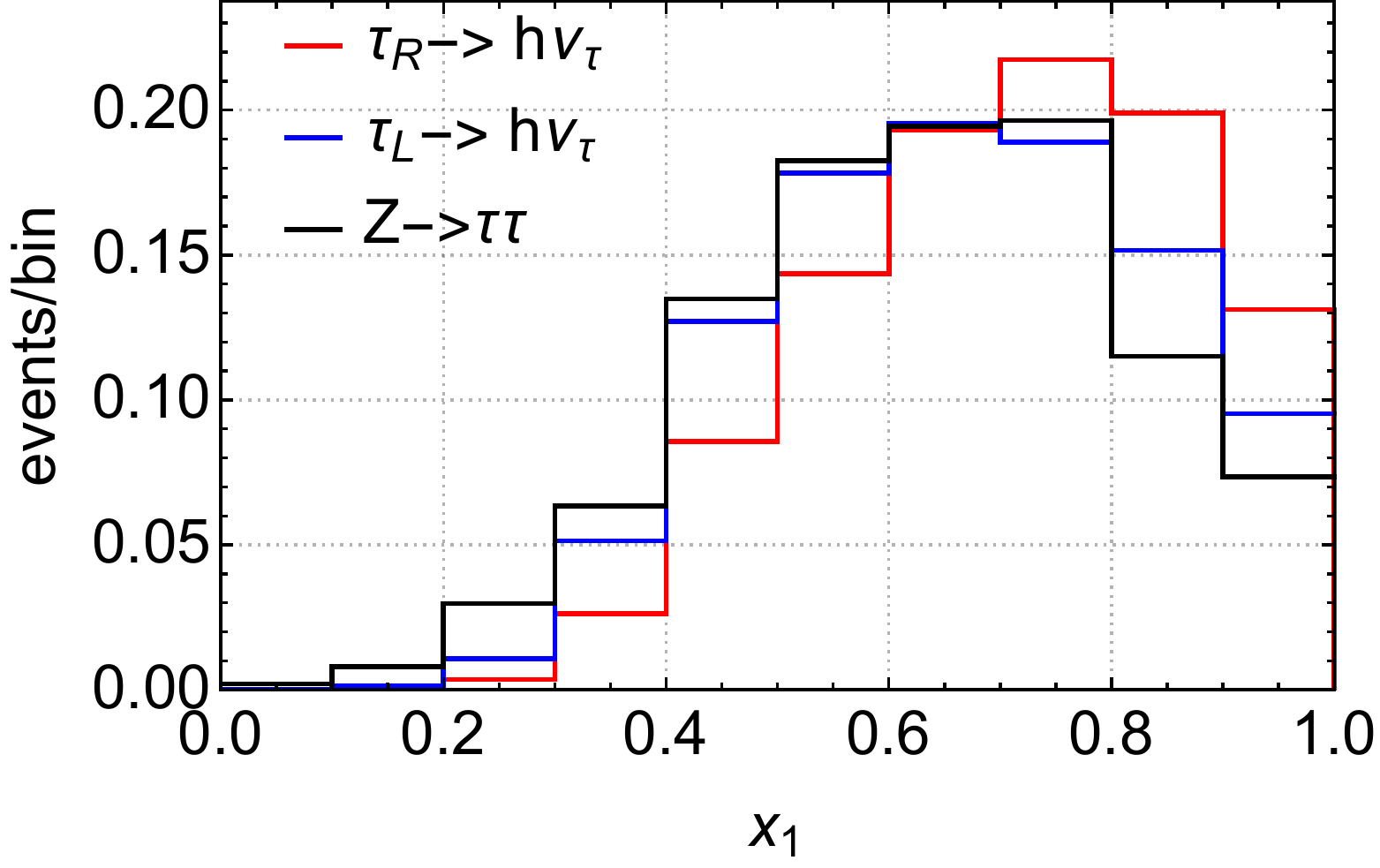}
\caption{
Reconstructed $x_1$ distributions for the signal $\tau_R$ (red) and $\tau_L$ (blue),
and for the $Z\to \tau\tau$ BG (black)  in the signal region with $\Delta m_{\rm col1}^{\rm th} = 25$~GeV. 
All the distributions are normalized to unity.
}
\label{fig:x1rec}
\end{figure}

In those three situations, we estimate how much we can constrain 
the parameters in ($y_{\mu\tau}, y_{\tau\mu}$) plane. 
As an illustration, the normalized reconstructed $x_1$ distributions for $\tau_R$, $\tau_L$, and $Z\to\tau\tau$ background are given in Fig. \ref{fig:x1rec}. 

For simplicity we consider the simplest two bin analysis 
by further dividing the signal regions $\Delta m_{\rm col1}^{\rm th} = 25$~GeV 
based on the reconstructed $x_1$ values 
into two signal regions SR$_1$ and SR$_2$.
We separate them as SR = SR$_1$$(x_1<0.6)$ + SR$_2$$(x_1\ge 0.6)$, 
and the corresponding number of events found in the SRs we denote as 
$N = N(x_1 < 0.6) + N(x_1 \ge 0.6) = N_1 + N_2$. 
We denote the signal and background contributions found in SR$_i$ ($i=1,2$)
as $S_i$ and $B_i$, respectively.

\begin{table}[h!]
\centering
\begin{tabular}{cc|cc|c|cc|c||ccc}
\hline
$\Delta m_{\rm col1}^{\rm th} $ &
SR &\multicolumn{3}{c|}{$N_{i, BR=0.12\%}$} & \multicolumn{3}{c||}{$N_i/N$} & \multicolumn{3}{c}{ $N_{i,{\rm obs}}$ for each scenario}\\
&&$\tau_R$& $\tau_L$ & $Z\to \tau\tau$ &$\tau_R$& $\tau_L$ & $Z\to \tau\tau$ &$\tau_R$& $\tau_0$& $\tau_L$ \\
\hline
\hline
\multirow{3}{*}{25~GeV} 
&SR$_1$   &   50.6 & 66.1 & 1692 & 0.26 & 0.37 & 0.42  & 3436 & 3443 & 3451 \\
&SR$_2$  &144.5 & 113.1 & 2331& 0.74 & 0.63 & 0.58  & 4807 & 4791 & 4776 \\
\cline{2-11}
&total                                  & 195.1 & 179.2 & 4023 &1&1&1& 8243 & 8234 & 8227\\
\hline
\hline
\multirow{3}{*}{5~GeV} 
&SR$_1$  &   17.8 & 25.6 & 136 & 0.24 & 0.37 & 0.37  & 289.8 & 293.7 & 297.6 \\
&SR$_2$  &56.2 & 43.6 & 80& 0.76 & 0.63 & 0.63  & 216.2 & 209.9 & 203.6 \\
\cline{2-11}
& total                                  & 74.0 & 69.2 & 216.0 &1&1&1& 506 & 503.6 & 501.2 \\
\hline
\end{tabular}
\caption{
Number of events in the SR$_1$ $(x_1 < 0.6)$ and SR$_2$ $(x_1\ge 0.6$) for ${\cal L}=36.1$~fb$^{-1}$. 
The results of the two possible choices for 
the width of the signal region $\Delta m_{\rm col1}^{\rm th} =25$ and 5~GeV are shown.
}
\label{tab:2dinfo}
\end{table}

\begin{figure}[h]
\centering
\includegraphics[width=0.32\textwidth]{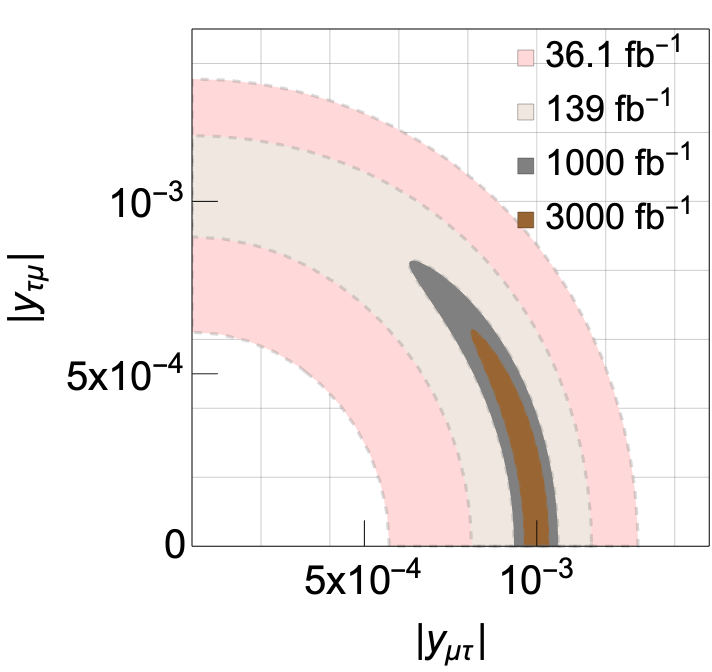}
\includegraphics[width=0.32\textwidth]{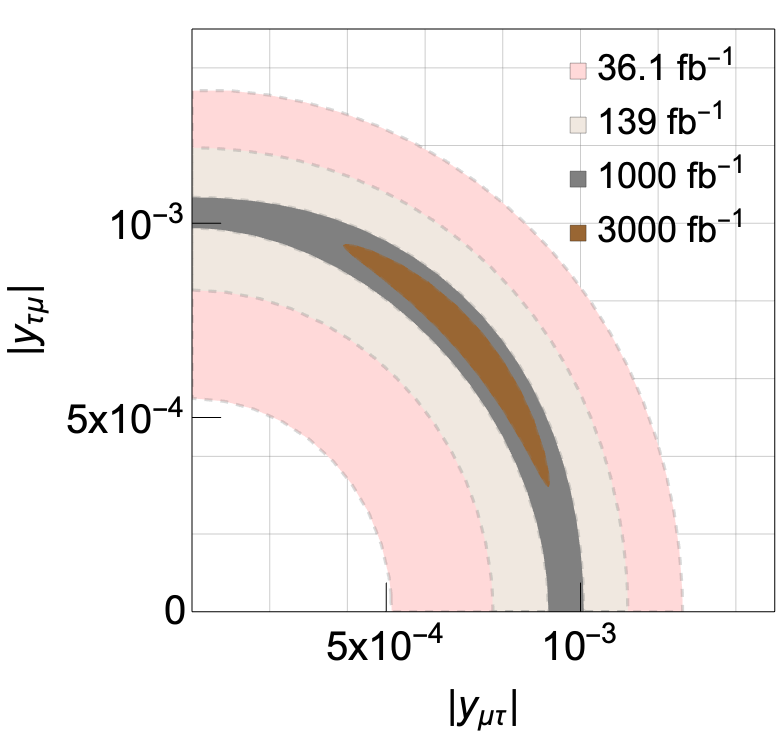}
\includegraphics[width=0.32\textwidth]{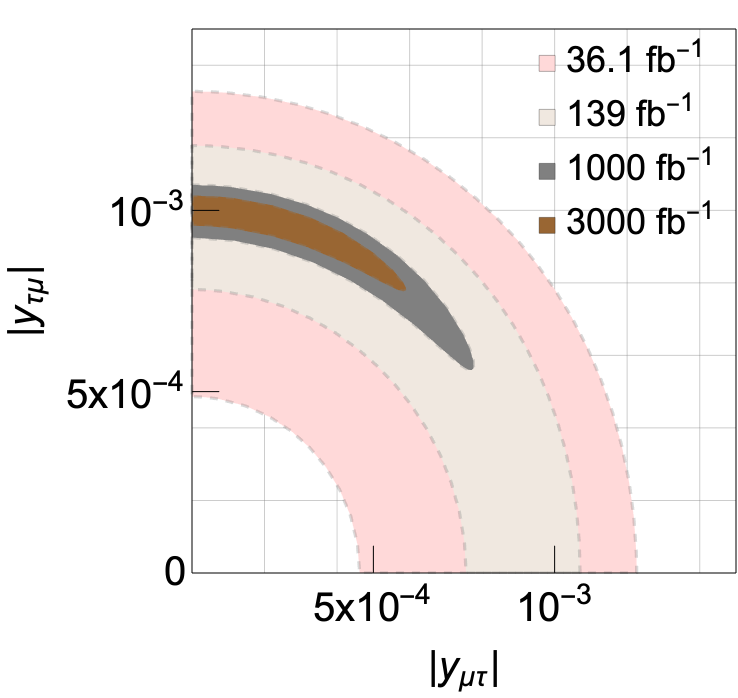}
\vspace{-0.2cm}
\caption{
Estimated sensitivity for the chirality structure in $h\to \tau\mu$ process
using the signal region with
$\Delta m_{\rm col1}^{\rm th}=25$~GeV.
The results for the three types of benchmark points predicting ${\rm BR}( h\to \tau\mu) = 0.12~\%$, 
$\tau_R$ scenario (left), $\tau_0$ scenario (center), and $\tau_L$ scenario (right) are shown. 
The $1\sigma$ contours for the integrated luminosity at 36.1, 139, 1000, and 3000~fb$^{-1}$ are shown.
}
\label{fig:contours}
\end{figure}

We consider the three scenarios mentioned above and assume the observed number of events for the two signal regions 
are given by $N_{i, {\rm obs}} = S_i + B_i$, where we assume $B_i = 2B_{i, Z}$ as before. 
From the two numbers $N_{1,{\rm obs}}$ and $N_{2,{\rm obs}}$, we can fit simultaneously $N_{R}$ and $N_{L}$.
The estimated signal numbers for each scenario at the integrated luminosity of  ${\cal L}=36.1$~fb$^{-1}$ 
are summarized in Table~\ref{tab:2dinfo}.
Based on the table, taking only statistical errors into account, a 1 $\sigma$ contour
is obtained as an ellipse in the
($\Delta N_R, \Delta N_L$) plane as follows:
\begin{align}
\chi^2 = \left( \frac{r\Delta N_R + l\Delta N_L}{\sqrt{N_{1, {\rm obs}}}} \right)^2
+ \left( \frac{(1-r)\Delta N_R + (1-l)\Delta N_L}{\sqrt{N_{2,{\rm obs}}}} \right)^2 \le 2.3,
\end{align}
where $\Delta N_R$ and $\Delta N_L$ are the deviations from the best fit values, 
which are the assumed numbers for $N_R$ and $N_L$ in each scenario. 
Depending on the scenario, we take 
$(N_{1,{\rm obs}}, N_{2,{\rm obs}}) = \{ (3435, 4807),(3443, 4791),(3451, 4776) \}\times \left({\cal L}/{\rm 36.1fb}^{-1}\right)$.
The parameters $r$ and $l$ are the the probabilities to fall into SR$_1$ 
for $\tau_R$ and $\tau_L$, respectively. We obtain them as $r=0.26$ and $l=0.37$ from Table~\ref{tab:2dinfo}.
The contours described by the Yukawa parameters are obtained using the following relationship, with $N_R = 195.1$, $N_L=179.2$, 
\begin{align}
&\Delta N_R = N_R \left( \frac{\cal L}{{\rm 36.1fb}^{-1}}\right) \frac{|y_{\mu\tau}|^2 - |\hat{y}_{\mu\tau}|^2}{10^{-6}}, \\ 
&\Delta N_L = N_L \left( \frac{\cal L}{{\rm 36.1fb}^{-1}}  \right) \frac{|y_{\tau\mu}|^2 - |\hat{y}_{\tau\mu}|^2}{10^{-6}}.
\end{align}
The expected $1\sigma$ contours for  $\Delta m_{\rm col1}^{\rm th}=25$~GeV are shown in Fig.~\ref{fig:contours}. 
We show the results for the integrated luminosity at 36.1, 139, 1000, and 3000~fb$^{-1}$.
For the conservative choice for the signal region width of  $\Delta m_{\rm col1}^{\rm th} =25$~GeV, 
the chirality structure would become sensitive after 1000~fb$^{-1}$. 
For example, extreme cases between the $\tau_R$ scenario and $\tau_L$ scenario 
can be distinguished at 2.3$\sigma$ (4.4$\sigma$) at 1000 (3000)~fb$^{-1}$. The $\tau_R$ scenario and $\tau_0$ scenario can be 
distinguished at 1.9$\sigma$ at 3000~fb$^{-1}$.
The reason that the sensitivity is not strong is due to the relatively large background contribution, 
which dilutes the sensitivity to the chirality of the system.

\begin{figure}[h!]
\centering
\includegraphics[width=0.32\textwidth]{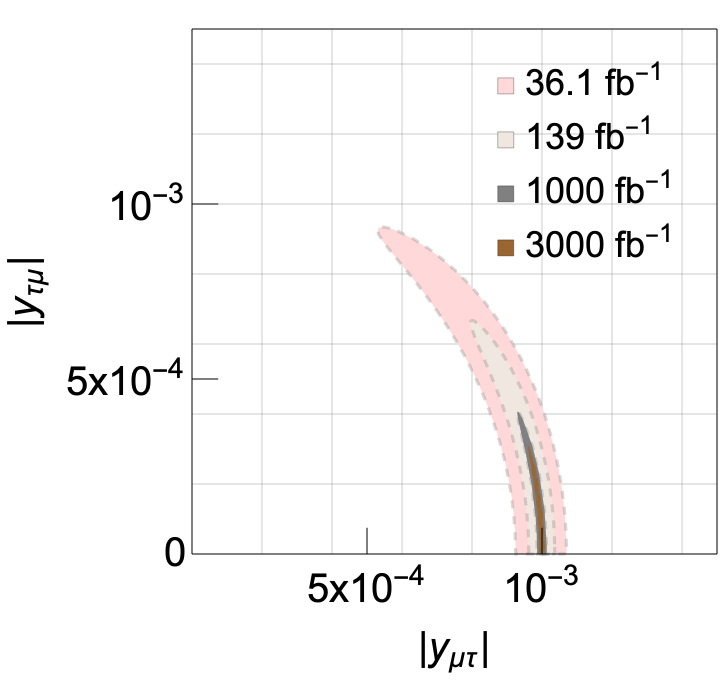}
\includegraphics[width=0.32\textwidth]{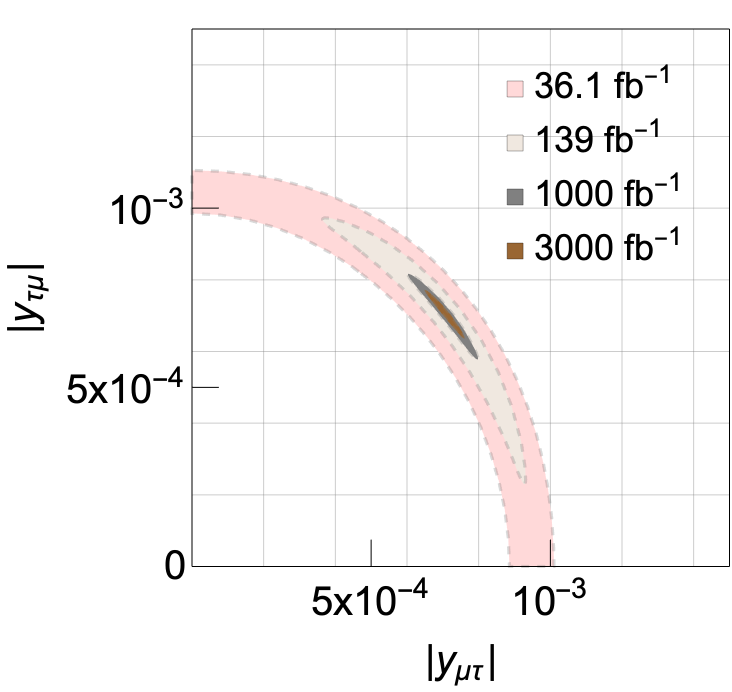}
\includegraphics[width=0.32\textwidth]{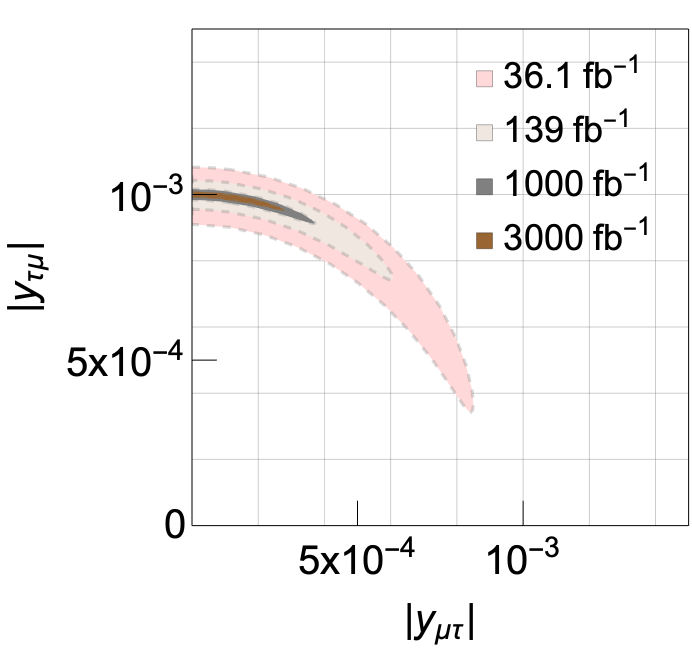}
\vspace{-0.2cm}
\caption{
Estimated sensitivity for the chirality structure in $h\to \tau\mu$ process using the signal region with 
$\Delta m_{\rm col1}^{\rm th}=5$~GeV.
The results for the three types of benchmark points predicting ${\rm BR}( h\to \tau\mu) = 0.12~\%$, 
$\tau_R$ scenario (left), $\tau_0$ scenario (center), and $\tau_L$ scenario (right) are shown. 
The $1\sigma$ contours for the integrated luminosity at 36.1, 139, 1000, and 3000~fb$^{-1}$ are shown.
}
\label{fig:contours5GeV}
\end{figure}

Further, we show the expected $1\sigma$ contours for $\Delta m_{\rm col1}^{\rm th}=5$~GeV in Fig.~\ref{fig:contours5GeV}
for the integrated luminosity at 36.1, 139, 1000, and 3000~fb$^{-1}$.
By taking the narrower signal region width of  $\Delta m_{\rm col1}^{\rm th} =5$~GeV, 
the signal over background ratio would become improved from about $1/40$ to about $1/6$, 
and the chirality structure would be distinguishable already at 139~fb$^{-1}$.
For example,
an extreme case $\tau_R$ scenario would be distinguished from the 
nonpolarized scenario ($\tau_L$ scenario) at 2.1$\sigma$ (4.8$\sigma$) level at 139~fb$^{-1}$.
Note that as we do not take the systematic uncertainty into account, these numbers 
should be rather optimistic and taken as reference values.
Nevertheless, we expect that accumulating more data would make such a sensitivity achievable and 
more detailed experimental study is desirable.

\section{Conclusion}
\label{sec:conclusion}
The Higgs LFV process is a smoking gun signature of new physics beyond the SM.
The chirality structure of the process is important information to discriminate the models,
but it is not often discussed in detail. Most of the experimental 
results reported basically assume no chirality preference.
In this paper, we consider how much we can probe the chirality structure in the $h\to \mu\tau$ process at the LHC.

We first discuss that to reconstruct the $h\to \tau\mu$ process, 
the collinear approximation with one missing particle assumption would be more effective than
that with two missing particle assumption. 
Thus, we compare the analysis with $m_{\rm col1}$ variable and that with $m_{\rm col2}$ variable.
We have shown that using the $m_{\rm col1}$ variable would improve
the signal over background ratio more easily than using the $m_{\rm col2}$ variable, 
since the $m_{\rm col1}$ distribution exhibits a sharp peak structure at the Higgs mass for the $h\to \tau\mu$ signal process.
We estimated the ultimate sensitivity of this process based on the $m_{\rm col1}$ analysis.
We then showed that the $\tau$ polarization affects the acceptance of the signature due to the jet $p_T$ threshold.
Consequently, the current search results should be altered by the polarization effects. 
We estimate the size of the effects and found that it is about 
at $\pm 4~\%$ level in terms of the BR$(h\to \tau\mu)$.
As a result the exclusion contour should become in general not a circle but an ellipse, 
where in general we have stronger constraints on $y_{\mu\tau}$ 
that governs the $h \to \tau_R\mu_L$ contributions.

Inspired by the recent 2$\sigma$ level  excess reported in this process,
we discuss whether the chirality structure is distinguishable or not, 
by considering the three benchmark scenarios with different chirality structures.
We utilize the reconstructed $x_1$ distributions for this purpose and 
demonstrate the simplest two bin analysis to obtain the 1$\sigma$ contours 
in two-dimensional parameter space for the several assumptions of the integrated luminosities.
Note that for this analysis, adopting $m_{\rm col1}$ analysis is important since 
appropriately estimating the invisible momentum of the $\tau$ decay is required to reconstruct the $x_1$ variable.
As a result, we found that the two extreme cases of the chirality structures, the $\tau_R$ scenario and $\tau_L$ scenario,
would be distinguishable at 1000~fb$^{-1}$ at $2\sigma$ level.
We also show that taking a narrower signal region increases the signal over background and enhances the sensitivity.
For this setup, we found the two extreme cases would be distinguishable already at 139~fb$^{-1}$
although a dedicated experimental study would be required to confirm the feasibility.
\medskip

Once we have a sensitivity for the chirality structure of the off-diagonal elements 
$y_{\mu\tau}$ and $y_{\tau\mu}$ separately, 
we would be able to distinguish the new physics models. 
For example, there are models predicting the following relation~\cite{Chiang:2015cba} in the 2HDM, 
\begin{equation}
\label{example}
{\cal L}_{\rm LFV} \propto \frac{m_\tau}{v} \tau_L \mu_R + \frac{m_\mu}{v} \mu_L \tau_R + h.c. . 
\end{equation}
Since $m_\tau \gg m_\mu$,
we can discuss whether these types of models are preferred or excluded.

Another interesting study to be done in the 2HDM framework would be to 
discuss the correlation between the hLFV and the chirality structure of the heavy resonances.
The off-diagonal $\xi_{\tau\mu}$ component contributes to the $h\to\tau \mu$ process and the couplings to the heavy resonances.
Thus, the existence of the hLFV process naturally predicts the existence of the LFV coupling to the heavy resonances, which
induces the LFV heavy resonance decay, for example, the heavy Higgs decay $H \to \tau\mu$. 
Following a similar analysis demonstrated in this paper, we would also be able to 
analyze the chirality structure of the Yukawa couplings to the heavy Higgses.
We leave this analysis for a future work.

%
%
%
%
%

\section*{Acknowledgements}

This work was supported, in part,
by the JSPS KAKENHI Grant, the Grant-in-Aid for Scientific Research\,A, No.~20H00160 (S. K., M. A.).
M.T. is supported by the Fundamental Research Funds for the Central Universities, 
the One Hundred Talent Program of Sun Yat-sen University, China,
and by the JSPS KAKENHI Grant, the Grant-in-Aid for Scientific Research\,C, Grant No.~18K03611.

\bibliography{biblio_arXiv}
\end{document}